\documentclass[a4paper,twocolumn]{aa}

\usepackage{txfonts}
\usepackage{natbib}

\bibpunct{(}{)}{;}{a}{}{,} 
\bibstyle{aa} 
\usepackage{rotating}

\usepackage{longtable}
\usepackage{lscape}
\usepackage{multirow}
\usepackage{dcolumn}
\usepackage{tabularx,calc}
\usepackage{graphicx}
\usepackage{graphics}

\begin{document}

\newcommand{\kev}{{\rm keV\,}}
\newcommand{\kevb}{{\rm {\bf keV\,}}}
\newcommand{\Gev}{{\rm GeV\,}}
\newcommand{\Mev}{{\rm MeV\,}}
\newcommand{\Mevb}{{\rm {\bf MeV\,}}}
\newcommand{\Mevd}{{\rm MeV}}
\newcommand{\Tev}{{\rm TeV\,}}
\newcommand{\beq}{\begin{equation}}
\newcommand{\eeq}{\end{equation}}

\newcommand{\cp}{}
\newcommand{\new}{}
\newcommand{\nnew}{}
\newcommand{\editor}{}
\newcommand{\etal}{et al.}

\def\agile {\emph{AGILE}}
\def\xmm {\emph{XMM-Newton}}
\def\cha {\emph{Chandra}}
\def\flux {\mbox{erg cm$^{-2}$ s$^{-1}$}}
\def\pflux {\mbox{ph cm$^{-2}$ s$^{-1}$}}
\def\lum {\mbox{erg s$^{-1}$}}
\def\nh {$N_{\rm H}$}
\def\sounum{54} %

\title{An updated list of AGILE bright $\gamma$--ray sources and their variability in pointing mode }

\author{F. ~Verrecchia$^{1,2}$, C. Pittori$^{1,2}$, A.W.~Chen$^{3}$, A.~Bulgarelli$^{4}$,
M. Tavani$^{5,6,7,8}$, F. Lucarelli$^{1,2}$, P.~Giommi$^{1,9}$,
S.~Vercellone$^{11}$, A.~Pellizzoni$^{10}$, A.~Giuliani$^{3}$, F.~Longo$^{12,13}$,
G.~Barbiellini$^{12,13,8}$,
M.~Trifoglio$^{4}$, F. ~Gianotti$^{4}$,
A.~Argan$^{5}$, L.A.~Antonelli$^{2,1}$,
P.~Caraveo$^{3}$, M. Cardillo$^{5,6}$, P.~W.~Cattaneo$^{14}$,
V.~Cocco$^{6}$,
S.~Colafrancesco$^{2,23}$, T.~Contessi$^{3}$,            
E.~Costa$^{5}$, E.~Del~Monte$^{5}$,
G.~De~Paris$^{5}$, G.~Di~Cocco$^{4}$, G.~Di~Persio$^{5}$,
I.~Donnarumma$^{5}$, Y.~Evangelista$^{5}$,
G.~Fanari$^{1}$,
M.~Feroci$^{5}$, A.~Ferrari$^{8,15}$,
M.~Fiorini$^{3}$, F.~Fornari$^{3}$, F.~Fuschino$^{4}$,
T.~Froysland$^{8,6}$, M.~Frutti$^{5}$,
M.~Galli$^{16}$, C.~Labanti$^{4}$, I.~Lapshov$^{5}$,
F.~Lazzarotto$^{5}$, F.~Liello$^{5}$, P.~Lipari$^{17,18}$,
E.~Mattaini$^{2}$, M.~Marisaldi$^{4}$, M.~Mastropietro$^{5,20}$, 
A.~Mauri$^{4}$, F.~Mauri$^{14}$, S.~Mereghetti$^{2}$,
E.~Morelli$^{4}$, E.~Moretti$^{7,8}$, A.~Morselli$^{7}$,
L.~Pacciani$^{5}$, F.~Perotti$^{3}$,
G.~Piano$^{5,8}$, P.~Picozza$^{6,7}$, M.~Pilia$^{22,10}$,  
C.~Pontoni$^{8,5}$, G.~Porrovecchio$^{5}$,
M.~Prest$^{21}$, R.~Primavera$^{1}$, G.~Pucella$^{19}$, 
M.~Rapisarda$^{19}$, A.~Rappoldi$^{14}$, E.~Rossi$^{4}$, A.~Rubini$^{5}$,
S.~Sabatini$^{10}$, P.~Santolamazza$^{1,2}$,
P.~Soffitta$^{5}$, S.~Stellato$^{1}$, E.~Striani$^{10}$,
F.~Tamburelli$^{1}$, A.~Traci$^{4}$, A.~Trois$^{10}$, E.~Vallazza$^{12,13}$,
V.~Vittorini$^{5,6}$, D.~Zanello$^{17,18}$,
 L.~Salotti$^{9}$ and G.~Valentini$^{9}$ }

\institute{ \centering \em{Affiliations can be found after the references}}


\offprints{F. Verrecchia, \\ \email{francesco.verrecchia@asdc.asi.it} }
\date{Received 12 March 2013; Accepted 14 jun 2013}
\authorrunning {F. Verrecchia \etal}

\titlerunning {An updated list of AGILE bright $\gamma$--ray sources and their variability in pointing mode}


\abstract{}
{We present a variability study of a sample of bright $\gamma$-ray (30\,\Mev\ -- 50\,\Gev) sources. This sample is an extension of the first AGILE catalogue of $\gamma$-ray sources (1AGL), obtained using the complete set of AGILE observations in pointing mode performed during a 2.3 year period from July 9, 2007 until October 30, 2009.}
{The dataset of AGILE pointed observations covers a long time interval and its $\gamma$-ray data archive is useful for monitoring studies of medium-to-high brightness $\gamma$-ray sources.
In the analysis reported here, we used data obtained with an improved event filter that covers a wider field of view, on a much larger (about 27.5 months) dataset, integrating data on observation block time scales, which mostly range from a few days to thirty days.
}
{ The data processing resulted in a better characterized source list than 1AGL was, and includes \sounum\ sources, 7 of which are new high galactic latitude ($|BII|\geq 5$) sources, 8 are new sources on the galactic plane, and 20 sources from the previous catalogue with revised positions. Eight 1AGL sources (2 high-latitude and 6 on the galactic plane) were not detected in the final processing either because of low OB exposure and/or due to their position in complex galactic regions. We report the results in a catalogue of all the detections obtained in each single OB, including the variability results for each of these sources. In particular, we found that 12 sources out of 42 or 11 out of 53 are variable, depending on the variability index used, where 42 and 53 are the number of sources for which these indices could be calculated. Seven of the 11 variable sources are blazars, the others are \object{Crab} pulsar+nebula, \object{LS I +61$^{\circ}$303}, \object{Cyg X-3}, and \object{1AGLR J2021+4030}.}
{}
\maketitle


\section{Introduction}
\label{In}

 The Astrorivelatore Gamma ad Immagini LEggero (AGILE) \citep{Tavani,Tavani1} is a mission of the Italian Space Agency (ASI) dedicated 
to $\gamma$-ray and hard X-ray astrophysics in the 30 \Mev-- 50\,\Gev\ and 18 -- 60\,\kev\ energy ranges, launched on April 23 2007.

AGILE is the first $\gamma$-ray mission to operate in space after the end of the EGRET \citep{egret} observations, and has been 
the only mission entirely dedicated to high-energy astrophysics above 30 \Mevd\ until 2008.
 On 11 June 2008, the Fermi Gamma-Ray Space Telescope \citep{GLAST,LAT} was launched, which is operating concurrently with AGILE.
The AGILE spacecraft had operated in fixed-pointing mode since launch until October 2009, completing 101 pointings or observation blocks (OBs). 
Then, due to a failure of the spacecraft reaction wheel, the attitude control system was reconfigured and the scientific mode of operation 
was changed to spinning mode. Currently, the instrument pointing direction scans the sky with an angular velocity of about 0.8$^{\circ}/$s, 
resulting in an exposure of about  $7\times 10^{6}$ cm$^{2}$s for about 70\% of the sky in one day.
 The best $\gamma$-ray sensitivity of AGILE is in the 100\,\Mev\ -- 1\,\Gev\ energy band.

The first AGILE catalogue of high-confidence $\gamma$-ray sources \citep[1AGLs;][]{1AGL} included a significance-limited (4 $\sigma$) sample of 47 sources, obtained with a conservative data analysis of the first-year inhomogeneous AGILE dataset.
 The second Fermi catalogue \citep[2FGLs;][]{Nolan12} increased the number of known \mbox{$\gamma$-ray} sources to about 2000, with the established main classes of active galactic nuclei (AGN), pulsars (PSRs) and pulsar wind nebulae. Since the third EGRET source catalogue (3EG; Hartman et al. 1999), \mbox{$\gamma$-ray} source flux variability on a 15-day time scale has been an important criterion for selecting candidates for multi-wavelength campaigns \citep[for example,][]{1999Rom,Tor01,Par05}.
 2FGL reported variability indices for each source over a monthly time scale.

We decided to study the flux variations of a sample of about 60 of the brightest sources over the OB time scale in the pointing mode phase, when the AGILE-GRID sensitivity was higher (the exposure efficiency per single source was 0.6 while that of Fermi/LAT is about 0.16, see Morselli et al. 2013).

\begin{figure}[ht]
\hbox{
\hskip -0.4truecm

\includegraphics[width=0.50\textwidth]{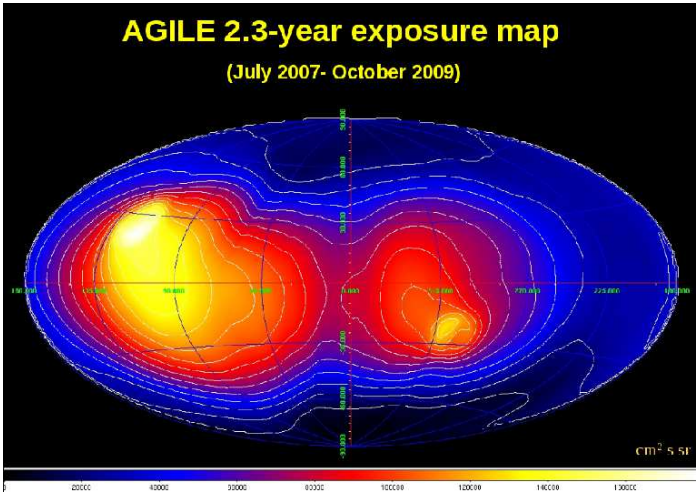}
}
\caption{ AGILE-GRID 2.3 year all-sky exposure map in an Aitoff projection obtained with the {\em FM} event filter from all pointed-observations data (96 OBs).
}

\label{fig:Fig1b}
\end{figure}

In this paper we present the results of the variability study of a sample of \sounum\ sources selected as a revision of the 1AGL, separately analysing each OB included in the 2.3 year dataset of AGILE pointed observations \citep[see also][]{CATV11}. The source sample was obtained with a preliminary revision of the 1AGL sources on the maps obtained from the merging of data from the whole dataset ({\em deep} maps hereafter). We used a much larger dataset than for 1AGL and data from the latest event filter ({\em FM3.119}, also known as FM; Bulgarelli et al., 2009, 2010 and Bulgarelli et al., in prep.; Chen et al. 2011b, 2012 and Chen et al., 2013), which covers a wider field of view (FoV), which allowed us to obtain a more uniform sensitivity over most of the sky despite the pointing strategy choosen in that phase; the mean effective exposure is $\sim$\,1800 Ms (see Fig.~\ref{fig:Fig1b}).

We describe the AGILE-GRID performances and the OB data archive in Sections 2 and 3, the data analysis procedure in Section 4, and the variability analysis in Section 5. In Section 6 we present the results obtained, both the variability results and the catalogue of all {\em detections} in all OBs. We then discuss the results and report some final considerations in Section 7.

\section{AGILE-GRID characteristics and calibration}
\label{resp-diff}

The AGILE scientific payload consists of a silicon tracker imaging detector \citep[Barbiellini et al. 2001, 2002;][]{2003NIMPA.501..280P}, a CsI(Tl) mini-calorimeter detector (Labanti et al. 2006, 2009), and an anti-coincidence system \citep{2006NIMPA.556..228P}, which together form the Gamma-Ray Imaging Detector (GRID). It also features the X-ray coded mask silicon imaging detector SuperAGILE \citep{2007NIMPA.581..728F}.
GRID is sensitive to photon energies in the 30 \Mev-- 50 \Gev\ energy band, with a wide FoV (2.5 sr in pointing mode) and accurate timing (2 $\mu$s), positional, and attitude information (15$\arcmin$ location accuracy for $>$\,10\,$\sigma$ detection). 

The GRID calibration has been updated since the launch \citep{1AGL,bulgaST,BTF11,BTFSP12,chen12Cal}.
The GRID instrument energy-dependent point spread function (PSF), derived from in-flight calibrations,
has a \mbox{full-width} at half-maximum (FWHM) of approximately 3.5$^{\circ}$ at 100 MeV and 0.7$^{\circ}$ at 1\,\Gev. The effective area is about 500 $\rm cm^2$ at 300\,\Mevd, depending on the event filter. 
Both GRID PSF and effective area have a very good \mbox{off-axis} performance and are well calibrated up to $60^\circ$, showing very smooth variations with the angle relative to the instrument axis.
The version of the response matrices used in this analysis (I0010) was developed through the analysis of in-flight data to introduce flux corrections to improve the spectral response, taking into account the energy dispersion. They were created by comparing the fluxes obtained with the AGILE likelihood analysis of the Vela pulsar in each energy bin, with those expected from the fluxes and spectra reported in the first Fermi catalogue (Chen et al. 2012, 2013).

The {\em FM} filter has a FoV of $\sim 60^{\circ}$ radius, which is different from the {\em F4} filter used for the 1AGL, which was optimized only up to off-axis angles of $\sim 40^{\circ}$ radius.
%

The data analysis used the AGILE diffuse emission model \citep[][]{giuliani04} for diffuse $\gamma$-ray background as for the 1AGL, which substantially improves upon the previous one used by EGRET. It was obtained by using state-of-the-art maps of neutral hydrogen \citep{kalberla05} and CO \citep{dame01}.
\vskip 0.5truecm

\section{OB data archive}
\label{sec:program}

The AGILE baseline pointing plan during the pointing mode phase was defined before the beginning of each year's announcement of opportunity cycle to reach the goals that maximize the scientific output of the mission. The AGILE Data Center (ADC), part of the ASI Science Data Center (ASDC) in Frascati, Italy\footnote{{\it http://agile.asdc.asi.it}}, is responsible for data reduction and scientific processing to create the official standard OB level-2 (LV2) data archive and distribute it to guest observers (GOs) or, when data become public, to the scientific community. 

Raw data are routinely archived, transformed in FITS format \citep[][]{trifoglio}, and then processed using the official scientific data reduction and analysis software tasks \citep[][]{bulga,chen11b}, which are integrated into an automatic quick-look (QL) pipeline system. The QL procedures perform data analysis on 1-, 2-, 3-, and 7-day time scales, including an automatic source detection and alert system \citep[][]{2008ADASS.XXX...YYY}. The final AGILE-GRID archive is created at the ADC after executing the OB standard analysis pipeline system. Both pipelines execute the event filter (based on a Kalman filter technique) for track identification, event direction, and energy reconstruction \citep{arem,giulianik1}\footnote{More details on the ADC organization and tasks will be given in a future publication (Pittori et  al. in preparation).}. 
The official OB pointing mode archive was built after removing data corresponding to slews and occasional losses of fine-pointing attitude, and it includes LV2 products, the spacecraft (LOG) and event (EVT) files, and the level-3 (LV3) counts, exposure and diffuse model maps. The OB LV3 maps used in this analysis were created in the E\,$>$\,100 \Mev\ energy band, with a bin of 0.25$^{\circ}\,\times$\,0.25$^{\circ}$, selecting events within 60$^\circ$ of the mean OB pointing direction and excluding photons

\longtab{1}{
\LTcapwidth=10.2in
\begin{landscape}
\scriptsize

\begin{longtable}{lllll||lllll} 
\caption{ AGILE pointings in the period 9 July 2007 - 30 October 2009, corresponding to the 96 observation blocks (OB). Acronyms used in table are:\\ ToO = target of opportunity pointing, SA = SuperAGILE special pointing.}\\
\hline
 & & & &  &  & & & &  \\

Target Name & OB  & Starting RA, Dec & Starting LII, BII &  Start/End Time ~~~~~~--~~~~~~~~~~~ Duration & Target Name & OB  & Starting RA, Dec & Starting LII, BII &  Start/End Time ~~~~~~--~~~~~~~~~~~ Duration  \\
  & Number  &  J2000 (deg) & (deg) &  ~~~~~ (UTC) ~~~~~~~~~~~~~~~~~~~~~~~~~~~~~ (days) &  & Number  &  J2000 (deg) & (deg) &  ~~~~~ (UTC) ~~~~~~~~~~~~~~~~~~~~~~~~~~~~~ (days) \\

 & & & &  &  & & & &  \\
\hline
 & & & &  &  & & & &  \\
\endhead

3C279 Region  &    900      &  195.60  ,  -6.65  &  307.81 ,  56.12 & 2007-07-09 12:00/07-07-13 12:00 ~~  4   & South Gal. Pole  &    5220  &   65.66  , -35.71  &  237.50 , -44.67 & 2008-02-12 12:00/08-02-14 12:00 ~~  2      \\
VELA Region   &    1000     &  157.98  , -60.21  &  286.42 ,  -1.90 & 2007-07-13 12:00/07-07-24 12:00 ~~ 11   &	Repointing       &     "    &           "        &          "       &      "          	              ~~  "       \\
ToO 3C 454.3  &    1100     &  17.83   ,  36.70  &  127.36 , -26.01 & 2007-07-24 12:00/07-07-30 12:00 ~~  6   & Musca Field      &    5300  &  191.93  , -71.89  &  302.64 ,  -9.02 & 2008-02-14 12:00/08-03-01 12:00 ~~	14   \\
ToO 3C 454.3  &    1150     &  17.83   ,  36.70  &  127.36 , -26.01 & 2007-07-24 12:00/07-07-30 12:00 ~~  6   & Gal. Center 1    &    5400  &  243.60  , -50.98  &  332.11 ,   0.02 & 2008-03-01 12:00/08-03-16 12:00 ~~	15   \\
VELA Region   &    1200     &  150.84  , -70.19  &  289.53 , -11.83 & 2007-07-30 12:00/07-08-01 12:00 ~~  2   & Gal. Center 2    &    5450  &  265.78  , -28.63  &  359.98 ,   0.63 & 2008-03-16 12:00/08-03-30 12:00 ~~	14   \\
SA Crab -45   &    1300     &  37.10   , 12.71   &  156.59 , -43.73 & 2007-08-01 12:00/07-08-02 12:00 ~~  1   & Anti-Center 1    &    5500  &  100.94  ,  21.71  &  192.64 ,   8.11 & 2008-03-30 12:00/08-04-05 12:00 ~~	 6   \\
VELA Region   &    1400     &  176.01  , -66.06  &  296.16 ,  -4.08 & 2007-08-02 12:00/07-08-12 12:00 ~~ 10   & SA Crab (8,24)   &    5510  &  108.28  ,  28.63  &  188.96 ,  17.00 & 2008-04-05 12:00/08-04-07 12:00 ~~	 2   \\
SA Crab -35   &    1500     &  47.41    , 16.08  &  164.83 , -35.32 & 2007-08-12 12:00/07-08-13 12:00 ~~  1   & SA Crab (15,26)  &    5520  &  111.76  ,  35.69  &  183.01 ,  22.20 & 2008-04-07 12:00/08-04-08 12:00 ~~	 1   \\
VELA Region   &    1600     &  195.55  , -66.56  &  304.00 ,  -3.72 & 2007-08-13 12:00/07-08-22 12:00 ~~  9   & Anti-Center 2    &    5530  &  110.40  ,  20.76  &  197.30 ,  15.72 & 2008-04-08 12:00/08-04-10 12:00 ~~	 2   \\
SA Crab -25   &    1700     &  57.14   , 18.57   &  171.08 , -27.31 & 2007-08-22 12:00/07-08-23 12:00 ~~  1   & Vulpecula Field  &    5600  &  286.26  ,  20.82  &   53.04 ,   6.47 & 2008-04-10 12:00/08-04-30 12:00 ~~	20   \\
VELA Region   &    1800     &  216.98  , -64.44  &  313.11 ,  -3.49 & 2007-08-23 12:00/07-08-27 12:00 ~~  4   & North Gal. Pole  &    5700  &  250.08  ,  72.50  &  104.85 ,  35.44 & 2008-04-30 12:00/08-05-10 12:00 ~~	10   \\
Galactic Plane   &    1900  &  236.57  , -41.87  &  334.44 ,  10.06 & 2007-08-27 12:00/07-09-01 12:00 ~~  4   & Cygnus Field 2   &    5800  &  304.29  ,  35.97  &   74.05 ,   0.27 & 2008-05-10 12:00/08-06-09 18:00 ~~	30.1 \\
SA Crab (15,15)  &    2000  &  69.48  , 5.59     &  190.90 , -26.29 & 2007-09-01 12:00/07-09-02 12:00 ~~  1   & ToO WComae/      &    5810  &  182.29  ,  29.61  &  195.50 ,  80.37 & 2008-06-09 18:00/08-06-15 12:00 ~~	 5.9 \\
SA Crab (0,15)   &    2100  &  68.21   , 20.57   &  177.13 , -18.28 & 2007-09-02 12:00/07-09-03 12:00 ~~  1   & ON+231          &     "    &          "           &           "          &       "                   ~~	"    \\	       
SA Crab (-15,15) &    2200  &  66.65   , 35.56   &  164.63 ,  -9.35 & 2007-09-03 12:00/07-09-04 12:00 ~~  1   & Cygnus Repointing&    5820  &  323.25  ,  50.08  &   93.66 ,  -1.17 & 2008-06-15 12:00/08-06-30 12:00 ~~	15   \\
Field 8          &    2300  &  51.41   , 71.02   &  134.88 ,  11.82 & 2007-09-04 12:00/07-09-12 12:00 ~~  8   & Antlia Field     &    5900  &  161.83 ,  -47.73  &  282.31 ,  10.11 & 2008-06-30 12:00/08-07-25 18:00 ~~	25   \\
SA Crab (0,5)    &    2400  &  78.54   , 21.73   &  182.16 ,  -9.89 & 2007-09-12 12:00/07-09-13 12:00 ~~  1   & TOO 3C 454.3     &    5910  &   19.37 ,  38.09   &  128.56 , -24.49 & 2008-07-25 18:00/08-07-31 12:00 ~~	 5.9 \\
Field 8          &    2500  &  74.88   , 58.33   &  150.99 ,   9.73 & 2007-09-13 12:00/07-09-15 12:00 ~~  2   & Estensione TOO   &    5920  &   25.09 ,  40.12   &  330.46 ,  28.98 & 2008-07-31 12:00/08-08-15 12:00 ~~	15   \\
SA Crab (45,0)   &    2600  &  84.21   , -23.01  &  226.70 , -26.12 & 2007-09-15 12:00/07-09-16 12:00 ~~  1   & 3C454.3          &     "    &           "        &           "          &       "                   ~~	"    \\	       
SA Crab (5,0)    &    2700  &  82.99   , 16.98   &  188.52 ,  -8.98 & 2007-09-16 12:00/07-09-17 12:00 ~~  1   & Musca Field 2    &    6010  &  175.31 ,  -74.13  &  298.10 , -11.92 & 2008-08-15 12:00/08-08-31 12:00 ~~	16   \\
SA Crab (0,0)    &    2800  &  83.77   , 22.03   &  184.62 ,  -5.67 & 2007-09-17 12:00/07-09-18 12:00 ~~  1   & ToO SGR 0501+4516&    6110  &   61.87 ,   44.06  &  333.90 ,  27.26 & 2008-08-31 12:00/08-09-10 12:00 ~~	10   \\
SA Crab (-5,0)   &    2900  &  84.62    , 27.05  &  180.77 ,  -2.33 & 2007-09-18 12:00/07-09-19 12:00 ~~  1   & Gal. Center 3    &    6200  &  256.55 ,  -28.53  &  355.51 ,   7.40 & 2008-09-10 12:00/08-10-10 12:00 ~~	30   \\
SA Crab (-15,0)  &    3000  &  85.35   , 37.09   &  172.59 ,   3.52 & 2007-09-19 12:00/07-09-20 12:00 ~~  1   & ToO PKS 0537-441 &    6210  &   98.80 , -46.77   &  255.44 , -22.05 & 2008-10-10 12:00/08-10-17 12:00 ~~	 7   \\
SA Crab (-25,0)  &    3100  &  86.17   , 47.12   &  164.26 ,   9.22 & 2007-09-20 12:00/07-09-21 12:00 ~~  1   & Aquila Field     &    6310  &  290.97 ,  10.10   &   45.62 ,  -2.51 & 2008-10-17 12:00/08-10-31 12:00 ~~	14   \\
SA Crab (-35,0)  &    3200  &  87.14   , 57.13   &  155.61 ,  14.60 & 2007-09-21 12:00/07-09-22 12:00 ~~  1   & Cygnus Field 3   &    6400  &   295.52 ,  35.64  &   70.03 ,   6.15 & 2008-10-31 12:00/08-11-30 12:00 ~~	30   \\
SA Crab (-45,0)  &    3300  &  88.35   , 67.14   &  146.45 ,  19.48 & 2007-09-22 12:00/07-09-23 12:00 ~~  1   & Cygnus Field 4   &    6500  &   320.40 ,  35.50  &   81.95 , -10.17 & 2008-11-30 12:00/08-12-20 12:00 ~~	20   \\
SA Crab (0,-5)   &    3400  &  90.10   , 22.14   &  187.54 ,  -0.59 & 2007-09-23 12:00/07-09-24 12:00 ~~  1   & Cygnus Field 5   &    6600  &   334.10 ,  44.05  &   95.70 , -10.47 & 2008-12-20 12:00/09-01-12 18:00 ~~	23.1 \\
SA Crab (15,0)   &    3500  &  91.03   , 7.14    &  201.11 ,  -7.14 & 2007-09-24 12:00/07-09-25 12:00 ~~  1   & ToO Carina Field &    6610  &   161.67 , -59.86  &  287.86 ,  -0.69 & 2009-01-12 18:00/09-01-19 18:00 ~~	 6.9 \\
SA Crab (25,0)   &    3600  &  91.84   , -2.88   &  210.46 , -11.12 & 2007-09-25 12:00/07-09-26 12:00 ~~  1   & Cygnus Field 6   &    6710  &   325.75 ,  68.11  &  106.75 ,  11.37 & 2009-01-19 18:00/09-02-28 12:00 ~~	 8.9 \\
SA Crab (35,0)   &    3700  &  92.50   , -12.93  &  220.02 , -14.95 & 2007-09-26 12:00/07-09-27 12:00 ~~  1   & Gal.Center 4     &    6800  &   247.20 , -29.03  &  349.85 ,  13.43 & 2009-02-28 12:00/09-03-25 12:00 ~~	25   \\
Crab Nebula      &    3800  &  94.32   , 22.05   &  189.52 ,   2.80 & 2007-09-27 12:00/07-10-01 12:00 ~~  4   & Gal.Center Prolonged & 6810 &   275.73 , -30.50  &    2.59 ,  -7.83 & 2009-03-25 12:00/09-03-31 12:00 ~~	 6   \\
SA Crab (0,-15)  &    3900  &  98.55   , 21.88   &  191.49 ,   6.19 & 2007-10-01 12:00/07-10-02 12:00 ~~  1   & Crab Field       &    6910  &   102.70 ,  31.71  &  184.07 ,  13.75 & 2009-03-31 12:00/09-04-07 12:00 ~~	 7   \\
SA Crab (-15,-15)&    4000  &  100.84  , 36.78   &  178.64 ,  14.35 & 2007-10-02 12:00/07-10-03 12:00 ~~  1   & Aquila Field 1   &    7010  &   288.88 , -19.31  &   18.06 , -13.82 & 2009-04-07 12:00/09-04-15 12:00 ~~	 8   \\
SA Crab (15,-15) &    4100  &  99.57   , 6.79    &  205.39 ,   0.18 & 2007-10-03 12:00/07-10-04 12:00 ~~  1   & Aquila Field 2   &    7100  &   290.88 ,  16.16  &   50.92 ,   0.44 & 2009-04-15 12:00/09-04-30 12:00 ~~	15   \\
Crab Field       &    4200  &  101.72  , 21.70   &  192.97 ,   8.76 & 2007-10-04 12:00/07-10-12 12:00 ~~  8   & Cygnus Field 7   &    7200  &   299.11 ,  29.78  &   66.49 ,   0.59 & 2009-04-30 12:00/09-05-15 12:00 ~~	15   \\
SA Crab (0,-25)  &    4300  &  110.13  , 20.72   &  197.23 ,  15.47 & 2007-10-12 12:00/07-10-13 12:00 ~~  1   & Vela Field 2     &    7300  &   127.35 , -37.14  &  256.41 ,   1.08 & 2009-05-15 12:00/09-05-25 18:00 ~~	10.1 \\
Gal. Center      &    4400  &  290.92  , -18.90  &   19.27 , -15.41 & 2007-10-13 12:00/07-10-22 12:00 ~~  9   & 3rd ToO 3C454.3  &    7310  &   328.44 ,  10.91  &   68.32 , -32.54 & 2009-05-25 18:00/09-05-29 12:00 ~~	 3.9 \\
SA Crab (0,-35)  &    4500  &  120.49  , 18.88   &  203.04 ,  23.74 & 2007-10-22 12:00/07-10-23 12:00 ~~  1   & Ripresa Vela Field 2&  7320 &   136.50 , -40.71  &  263.67 ,   4.38 & 2009-05-29 12:00/09-06-04 12:00 ~~	 6   \\
Gal. Center Reg. &    4600  &  301.17  , -17.11  &   25.10 , -23.67 & 2007-10-23 12:00/07-10-24 08:00 ~~  1   & Virgo Field 2    &    7410  &   167.14 ,  10.71  &  242.13 ,  60.76 & 2009-06-04 12:00/09-06-15 12:00 ~~	11   \\
ToO 0716+714     &    4610  &  148.94  , 67.89   &  143.36 ,  41.59 & 2007-10-24 08:00/07-10-29 12:00 ~~  5   & Cygnus Field 8   &    7500  &   330.22 ,  43.11  &   92.83 ,  -9.58 & 2009-06-15 12:00/09-06-25 12:00 ~~	10   \\
ToO Extended     &    4630  &  157.46  , 66.94   &  141.55 ,  44.72 & 2007-10-29 12:00/07-11-01 12:00 ~~  3   & Cygnus Field 9   &    7600  &   344.77 ,  37.90  &   99.66 , -19.84 & 2009-06-25 12:00/09-07-15 12:00 ~~	20   \\
SA Crab (0,-45)  &    4700  &  130.61  , 16.34   &  209.79 ,  31.74 & 2007-11-01 12:00/07-11-02 12:00 ~~  1   & Cygnus Field 10  &    7700  &   330.35 ,  64.26  &  105.73 ,   7.23 & 2009-07-15 12:00/09-08-12 12:00 ~~	28   \\
Cygnus Region    &    4800  &  296.88  , 34.50   &   69.59 ,   4.62 & 2007-11-02 12:00/07-12-01 12:00 ~~ 29   & Vela Field 3     &    7800  &   202.30 , -62.10  &  307.33 ,   0.45 & 2009-08-12 12:00/09-08-31 12:00 ~~	19   \\
Cygnus Field 1   &    4900  &  304.43  , 53.55   &   88.82 ,   9.93 & 2007-12-01 12:00/07-12-05 09:00 ~~  4   & Norma Field      &    7900  &   243.77 , -35.45  &  343.05 ,  11.11 & 2009-08-31 12:00/09-09-10 12:00 ~~	10   \\
Cygnus Repointing&    4910  &  322.50  , 38.24   &   85.12 ,  -9.42 & 2007-12-05 09:00/07-12-15 12:00 ~~ 10.1 & SA Crab (15,6)   &    8000  &    78.33 ,  6.66   &  195.06 , -18.31 & 2009-09-10 12:00/09-09-13 12:00 ~~	13   \\
Cygnus Repointing&    4920  &  322.50  , 38.24   &   85.12 ,  -9.42 & 2007-12-15 12:00/07-12-16 12:00 ~~  1   & SA Crab (25,3)   &    8100  &    81.78 , -3.12   &  205.88 , -20.15 & 2009-09-13 12:00/09-09-16 12:00 ~~	 3   \\
Virgo  Field     &    5010  &  173.43  , -0.44   &  265.65 ,  56.70 & 2007-12-16 12:00/08-01-08 12:00 ~~ 33   & Galactic Center 5&    8200  &   263.19 , -23.49  &  232.78 , -28.89 & 2009-09-16 12:00/09-09-30 12:00 ~~	14   \\
Vela  Field      &    5100  &  147.06  , -62.52  &  283.47 ,  -6.79 & 2008-01-08 12:00/08-02-01 12:00 ~~ 24   & Aquila Field 3   &    8300  &   278.13 , -23.22  &   10.11 ,  -6.45 & 2009-09-30 12:00/09-10-15 12:00 ~~	15   \\
South Gal. Pole  &    5200  &   58.35  , -37.80  &  240.39 , -50.58 & 2008-02-01 12:00/08-02-09 09:00 ~~  8   & Aquila Field 4   &    8400  &   285.78 ,  28.82  &   60.12 ,  10.38 & 2009-10-15 12:00/09-10-31 12:00 ~~	16   \\
ToO MKN 421      &    5210  &  250.97  ,  50.29  &   77.31 ,  40.63 & 2008-02-09 09:00/08-02-12 12:00 ~~  3   &  & & & &  \\

 & & & &  &  & & & &  \\
\hline
\label{tab:pointings}

\end{longtable}
\normalsize
\end{landscape}
}

\noindent within 80$^{\circ}$ from the reconstructed satellite-Earth vector, to reduce $\gamma$-ray Earth-albedo contamination. 
This archive was created using the latest official version of the standard software and instrument response functions available at the time of the analysis\footnote{Software build version: {\rm BUILD GRID\_STD\_17} and {\rm BUILD GRID\_SCI\_19}; calibration files version {\rm I0010}.}.
The complete pointing mode OB archive\footnote{Public OB archive is available at\\{\it http://www.asdc.asi.it/mmia/index.php?mission=agilemmia}.} is composed of 101 datasets with non-uniform exposures (from 1 to 45 days). We analysed the scientifically validated data archive that covers 2.3 years with 96 OBs (see Table \ref{tab:pointings}), not including the first 5 OBs of the commissioning phase.

\section{OB data analysis procedure}
\label{sec:obarch}

The procedure developed for the source analysis on the whole archive is based on point source detection at fixed preselected positions.
The AGILE multi-source analysis task \citep[ALIKE, ][]{bulga,chen11b}, which we used to derive the best estimates of point source parameters, i.e. source significance, $\gamma$-ray flux, Galactic coordinates, and 95\% confidence level (c.l.), is based on the maximum likelihood (ML) method.
The ML method, used in previous $\gamma$-ray data analysis \citep{1996ApJ...461..396M} and for all recent AGILE and Fermi imaging results \citep[see for instance][]{BulgCyg,BulgML,Nolan12}, compares measured counts in each pixel with the predicted counts derived from a $\gamma$-ray model composed of the Galactic diffuse model (see Section \ref{resp-diff}), an isotropic diffuse component, and point sources modelled according to the instrument PSF. To reduce systematic errors that arise because the data were taken far away from the source, only data within a circle of a given radius around each source were used.
 The significance of a source detection is given by the test statistic TS, defined as the likelihood ratio, that is expected to behave as $\chi$$^{2}_{1}$ in the null hypothesis for one single parameter that is to be optimized, i.e. the source flux. When $m$ parameters must be optimized, the TS behaves as $\chi$$^{2}_{m}$.\\

 We divide the procedure in three main steps:
\begin{itemize}
\item[1)] the preliminary reanalysis of the 1AGL catalogue, mainly in complex Galactic plane regions, based on LV3 deep maps from the merging of all data in the OB archive, more recent AGILE results and including some high-latitude ($|BII|\,>\,10^{\circ}$) sources detected in QL automatic procedures on a weekly time scale. The creation of the updated source list for next step;

\item[2)] the execution of a procedure based on the AGILE multi-source ML task on each OB LV3 map, first allowing source fluxes to vary while keeping source positions fixed, then allowing the source positions to vary;

\item[3)] the revision of the source list, taking into account the new results at fixed positions, for example when a source $\sqrt{TS}$ is significantly changed (increased or decreased). The 95\% c.l. contour obtained leaving the source position free was also checked to verify that contours of transient sources did not overlap the nearest source position.
\end{itemize}

\noindent where we used the $\sqrt{TS}$ defined above as detection significance parameter.

The reanalysis of the 1AGL source positioning in step 1) is described in Section \ref{sec:firstagcat}, followed by a description of step 2) in Section \ref{sec:obsub} and step 3) in Section \ref{sec:detselvar}.

\subsection{Revision of the First AGILE-GRID catalogue based on the 2.3 yr dataset}
\label{sec:firstagcat}

The first AGILE catalogue \citep{1AGL} was built using data collected during the first year of operations (July 2007 - June 2008), which includes half of cycle-1 of the AGILE GO program. The data analysis based on the conservative F4 filter, and the non-uniform sensitivity due to the inhomogeneous sky coverage during AGILE first year, limited the results in complex Galactic regions. We reanalyzed 1AGL sources in the Galactic plane regions, such as the Carina, LS I +61$^{\circ}$303, Cygnus, Crux, and near the Galactic center, using the updated E\,$>$\,100 \Mevd\ deep maps obtained from the merging of data up to October 2009 (Fig. \ref{fig:Fig1b}) processed with the {\em FM} event filter. These maps were created using the event selections described in Section \ref{sec:program}, using both 0.05$^{\circ}\,\times$\,0.05$^{\circ}$ and 0.1$^{\circ}\,\times$\,0.1$^{\circ}$ bins.
We then built an updated reference list with improved source positions in these regions. We use the abbreviation "1AGLR" for the repositioned and new sources. 
First, we reanalyzed the low-significance (3\,$\leq$\,$\sqrt{TS}$\,$\leq$\,4) detections from the 1AGL processing in the Carina (283$^{\circ}$\,$<$\,LII\,$<$\,292$^{\circ}$), LS I +61$^{\circ}$303 (130$^{\circ}$\,$<$\,LII\,$<$\,136$^{\circ}$) and Scutum-Sagittarius (6$^{\circ}$\,$<$\,LII\,$<$\,36$^{\circ}$) regions, as well as four high-latitude detections at $\sqrt{TS}$\,$\geq$\,4 not included in the catalogue (three of which are included in the final source list: \object{1AGLR J0135+4759}, \object{1AGLR J0222+4305}, \object{1AGLR J0321+4137}). We then reanalyzed sources from more recently refined AGILE results in the Galactic plane \citep[Tavani et al. 2009b, 2009c; ][]{cygx1,giulianiw28,BulgCyg,PianoCyg}. 

\begin{figure}[ht]
\begin{center}
\hskip -0.2truecm
\vbox{

\includegraphics[width=0.50\textwidth]{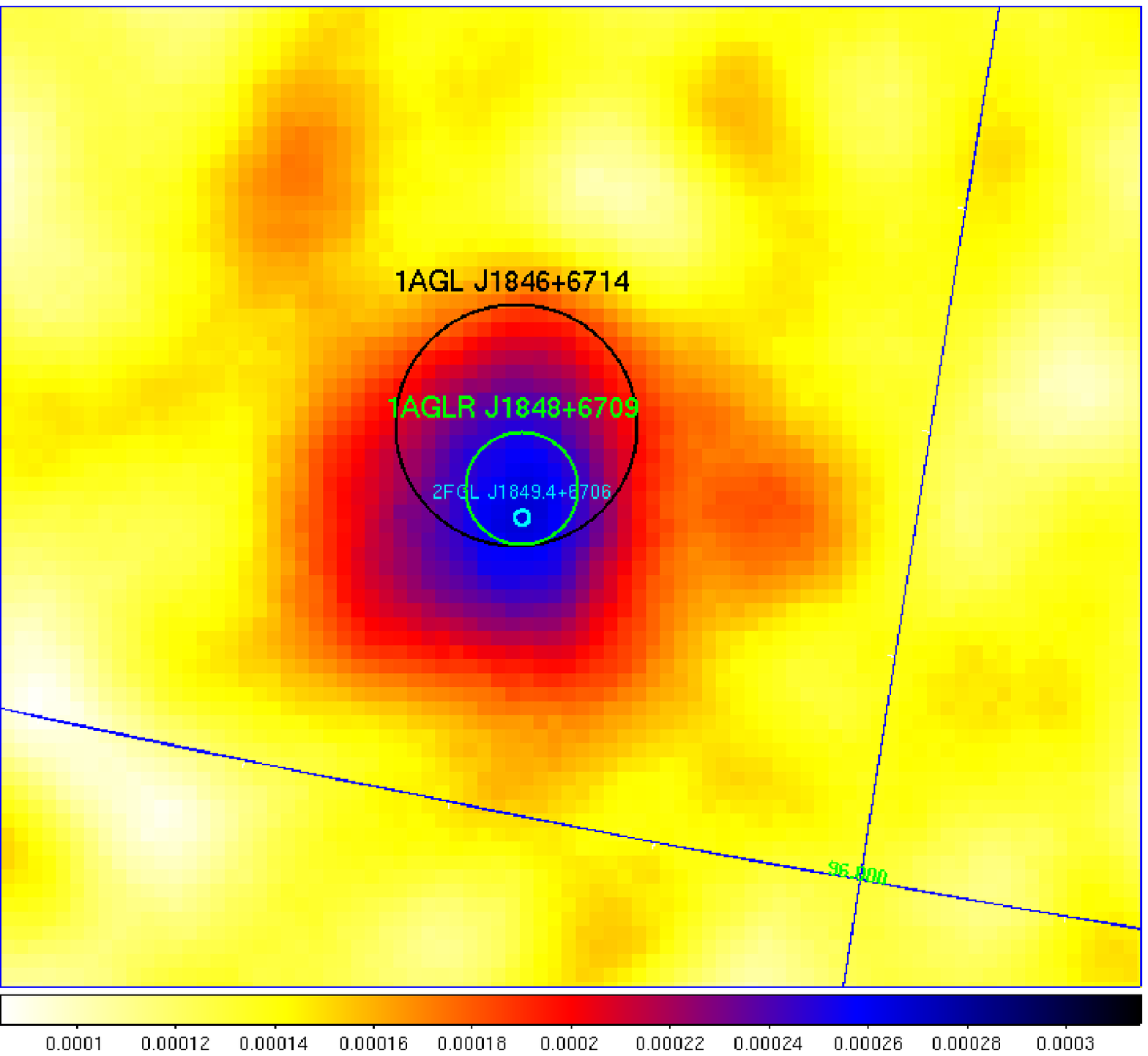}

\includegraphics[width=0.50\textwidth]{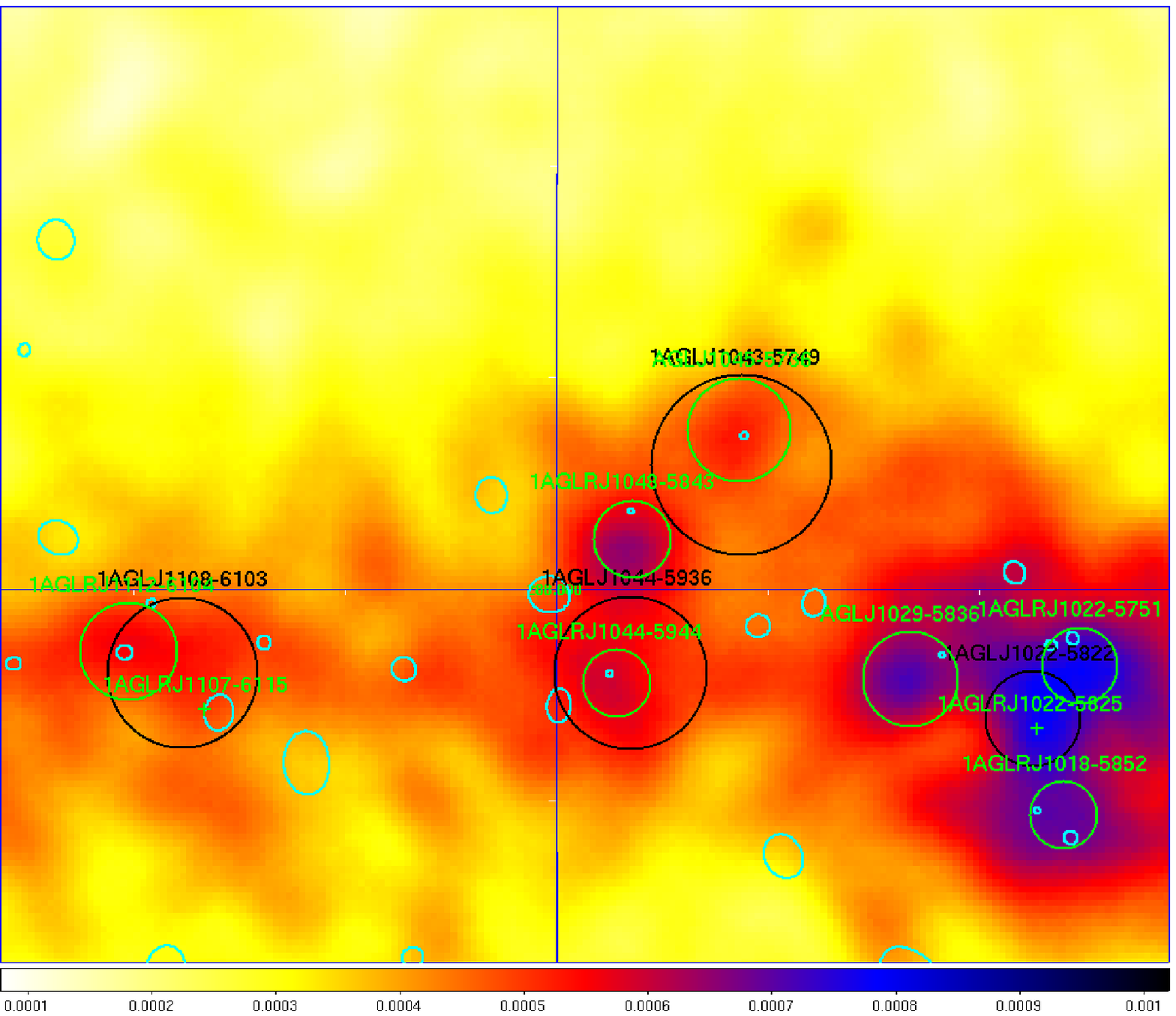}
}
\end{center}
\vskip -0.3truecm
\caption{Intensity maps of two 1AGL field refinements. In the first panel a simple case is shown, the high-latitude source \object{1AGL J1846+6714} repositioned to \object{1AGLR J1848+6709}. The new smaller error circle is indicated in green inside the 1AGL circle, while the \object{2FGL J1849.4+6706} error ellipse is plotted in cyan. In the second panel a complex case is shown, the Carina region. New sources and new positions for 1AGL sources are indicated with a green circle, except for two sources for which no 95\%\ c.l. contour has been obtained; their positions are indicated with crosses. The 1AGL error circles are indicated in black, the 2FGL error ellipses in cyan.
}
\label{fig:Fig2}
\end{figure}

We performed an ML multi-source analysis on the updated deep maps for these sources, to obtain the average fluxes used in the subsequent individual OB analyses. Two examples of source repositioning are shown in Fig. \ref{fig:Fig2}: a simple case for the high-latitude source \object{1AGLR J1848+6709} and a complex case for the Carina region. For this last case new source positions obtained in the ML analysis where source positions were allowed to vary, the 95\% c.l. contour radii for the sources for which they were closed and their $\sqrt{TS}_{deep}$ values are reported in Table \ref{tab:newgsou}. Two of these sources, \object{1AGLR J1022-5825} and \object{1AGLR J1107-6115}, have been included in the list although they do not have a 95\% c.l. contour from the ML analysis on deep maps. In these cases an additional $\sqrt{TS}_{deep}$ value is shown, marked with an asterisk ($\sqrt{TS}_{deep}^{*}$ hereafter); the value was obtained by keeping source position fixed.

 We included \object{1AGLR J1022-5825} because it is only slightly repositioned with respect to \object{1AGL J1022-5822}, whose error radius did not include the new sources (\object{1AGLR J1018-5852}, \object{1AGLR J1022-5751} and \object{AGL J1029-5836}), however, and also because it is significant on a single OB (7800) map. It was possible to obtain a detection at $\sqrt{TS}$\,$=$4.4 on this OB, with an acceptable 95\% c.l. contour, when allowing the source position to vary. Moreover, by executing an ML analysis on deep maps, keeping its position and the parameters of all nearby sources fixed, we obtained $\sqrt{TS}_{deep}$\,$=$\,4.0.
\object{1AGLR J1107-6115} was analogously included because of $\sqrt{TS}_{deep}^{*}$\,$\geq$\,4 obtained with the deep maps at fixed position, and $\sqrt{TS}$\,$\geq$\,4 in some OBs with an ML analysis with the parameters of all nearby sources (and of the diffuse emission) fixed.
 \object{AGL J1029-5836} and \object{AGL J1045-5736} (a repositioning of \object{1AGL J1043-5749}), which have a $\sqrt{TS}_{deep}$ above 5, were instead not significant over the OB time scale, and so do not appear in the final table.

\begin{table}[tbh]
\begin{center}
\begin{tabular}{|l l l l |}
\hline
\multicolumn{1}{|l}{Source Name} & 
\multicolumn{1}{l}{R.A., Dec.} & 
\multicolumn{1}{l}{95\% Error} & 
\multicolumn{1}{l|}{$\sqrt{TS}_{deep}$} \\[3 pt]
 &  & Radius  &  \\
 & ($^\circ$) & ($^\circ$) &   \\
\hline
& & &  \\
 \object{1AGLR J1018-5852} & 154.403, -58.860  & 0.25 & 10.5 \\[3 pt]
 \object{1AGLR J1022-5751} & 155.389, -57.846  & 0.29 &  7.4 \\[3 pt]
 \object{1AGLR J1022-5825} & 155.406, -58.423  &  --  & $<$2, 4.0$^{*}$ \\[3 pt]  
 \object{AGL J1029-5836}   & 157.351, -58.608  & 0.36 &  5.0 \\[3 pt]  
 \object{AGL J1045-5736}   & 161.253, -57.602  & 0.39 &  5.5 \\[3 pt]  
 \object{1AGLR J1044-5944} & 161.108, -59.734  & 0.25 &  4.9 \\[3 pt]  
 \object{1AGLR J1048-5843} & 161.887, -58.712  & 0.29 &  5.6 \\[3 pt]
 \object{1AGLR J1107-6115} & 166.631, -61.257  &  --  & $<$3, 4.1$^{*}$ \\[3 pt]
 \object{1AGLR J1112-6104} & 168.092, -61.073  & 0.44 &  5.4 \\[3 pt]  

\hline
\end{tabular}
\end{center}

\caption{Revised $\gamma$-ray source list in the Carina region. 95\% error radii are shown for the sources with closed contours. For \object{1AGLR J1022-5825} and \object{1AGLR J1107-6115}, an additional $\sqrt{TS}_{deep}$ value $^{*}$ is shown, obtained by keeping the source position fixed. } 
\label{tab:newgsou} 
\end{table}

We also included seven sources detected in the QL processing on a weekly time scale with $\sqrt{TS}$\,$>$\,4 at medium-high galactic latitude ($|BII|$\,$>$\,5$^\circ$) to verify their significance on the non-uniform OB time scale. Only four of these were detected in our final processing and are included in the final source list described in Section \ref{sec:results} (they are \object{1AGLR J1625-2531}, \object{1AGLR J1803-3941}, \object{1AGLR J1833-2057}, \object{1AGLR J2030-0617}). The results from the QL processing on various time scales will be reported elsewhere.

We compared the results of each run with the input reference source list, revising it if needed. We took into account the 95\% c.l. contours and the $\sqrt{TS}$, to check in detail candidate sources at fluxes\,$\sim$\,3.0\,$\times 10^{-7}$\,\pflux in the Carina, LS I +61$^{\circ}$303, Scutum and \object{1AGLR J1625-2531} regions. 
This review resulted in the final version of the list, which includes 90 candidate sources. This source list was used as input in the following step.

 A better characterization of complex Galactic regions will be reported in a future catalogue, implementing a new automatic source blind search.

\subsection{Analysis of individual OBs}\label{sec:obsub}

The ALIKE task is an iterative procedure that evaluates all the ML parameters that are left free to vary for each source in the input source list.
We decided to fix the positions of all sources to reduce the number of parameters evaluated by the ALIKE task during the ML analysis in each single OB. This reduces the task running-time and can improve the precision of the results \citep[see ][]{BulgML}. For the same reason, we also fixed the spectral indices to -2.1 for all sources except for the five well-known bright pulsars \object{Geminga}, Crab, \object{Vela}, \object{1AGL J1709-4428}, and \object{1AGLR J2021+3653} \citep[fixed to -\,1.66, -\,2.2, -\,1.69, -\,1.86 and -\,1.86, from 3EG; ][]{1999ApJS..123...79H,chen12Cal}, for \object{1AGLR J0007+7307}, \object{1AGLR J2031+4130} and \object{1AGL J1836+5923} (fixed to -\,1.85, -\,1.84 and -\,1.65, from 3EG and Bulgarelli et al. 2008, 2012a), for the high-mass X--ray binaries (HMXBs) LS I +61$^{\circ}$303 and Cyg X-3 \citep[fixed to -\,2.21 and -\,2.0 from 3EG and][]{BulgCyg} and the supernova remnants (SNRs) \object{W28}, \object{IC443} and \object{W44} \citep[fixed to -\,1.8, -\,2.01 and -\,1.93 from][and 3EG respectively]{giulianiw28}. These values are compatible with the 2FGL ones at the AGILE spectral resolution.
 In this analysis we used the ML task {\rm AG\_multi2} included in the AGILE {\rm BUILD GRID\_SCI\_20} package and a circle of radius 10$^{\circ}$ for the ML analysis.

To improve the reference list (in conjunction with the creation of a reprocessed OB archive with a new software version), in particular the multi-source detection in complex regions, all steps of the analysis were repeated five times.
Finally, we selected the results obtained with the last processing with source positions kept fixed to the last version of the reference source list (see Section \ref{sec:detselvar}).
Our goal was to refine multi-source detection in complex regions and add some high-latitude source, therefore we did not execute any automatic source detection process.

\section{Detection selection and variability analysis}
\label{sec:detselvar}

To investigate the variability within the OB dataset for established sources on deep maps and search for flaring episodes of sources with mean low significance, we analyzed the results from the last execution (the fifth) of the ML multi-source procedure on sources in the input reference list according to their significance and whether or not a 95\% location contour could be found. 
We chose sources with either 
\begin{itemize}
\item[1)] both a high-significance detection ($\sqrt{TS}\ge4$) from the deep map analysis and at least one OB detection with $\sqrt{TS}\ge 3$,
 or
\item[2)] at least one detection with $\sqrt{TS}\ge4$ in a single OB.
Moreover, we selected only sources with a 95\% location contour from the deep map analysis (with a few exceptions, see section \ref{sec:firstagcat}).
\end{itemize}
Then we checked each single OB detection with $3\leq\sqrt{TS}<4.5$, inspecting the single OB intensity map and checking the size and shape of the source 95\% c.l. contours from the ML at free position, if present, to verify whether it overlapped the nearest sources.
We thus excluded eight sources from the final source list. All sources in the last version of the reference list that do not appear in the catalogue table were not significant in our analysis according to these criteria.

We used the method developed by McLaughlin et al. (1996) that was recently used in the analysis of \object{1AGL J2022+4032} data \citep{chen11} to study the OB-scale $\gamma$-ray flux variability of the selected sources and report the results in the next section.
We computed three versions of the variability index. To calculate the traditional index $V$, we computed the weighted mean flux, its error, and the $\chi^{2}$ from the fluxes and errors in each OB. We evaluated the probability $Q$ of having a value of $\chi^{2}\geq\chi^{2}_{observed}$ for a source assumed to have constant flux, and then defined the variability index $V=-\log Q$. Sources with $V<0.5$ were classified as {\it non-variable}, with $0.5\leq\,V<1$ as {\it uncertain}, and with $V\geq1$ as {\it variable}. 
A high $V$ may indicate either strong flux variations or weak flux variations detected with small flux errors. Moreover, sources with a very small number of detections may also have low $V$ \citep{Nolan03}.
We computed a second version of the variability index, $V_{sys3\sigma}$, by adding a 10\% systematic component to each of the flux errors, where the 10\% figure is based on Monte Carlo simulations, pre-flight beam tests and in-flight analysis of known sources. Finally, we computed a third version, $V_{sys2\sigma}$, of the V parameter as $V_{sys3\sigma}$, which also included the flux measures with $2\leq\sqrt{TS}\leq3$, for which the ALIKE task produces a 95\% c.l. flux upper limit (UL). In this case we set the flux errors to $(F_{UL} - F_{i})/2$. The two indices, $V_{sys3\sigma}$ and $V_{sys2\sigma}$, which take systematic errors into account, will tend to be lower than $V$, which takes into account only the statistical error. These indices serve as the reference ones in our variability analysis, where $V_{sys2\sigma}$ allows a larger detections sample for each source to provide some variability information for sources with few pointings.
We excluded from the calculation of the indices the detections with a mean effective exposure lower than $9\times 10^{6}$ cm$^{2}$s, corresponding to a small number of detections in OBs with a few days duration executed during the verification phase (July--November 2007).

\begin{figure}[ht]
\hskip -0.3truecm
\includegraphics[width=0.53\textwidth]{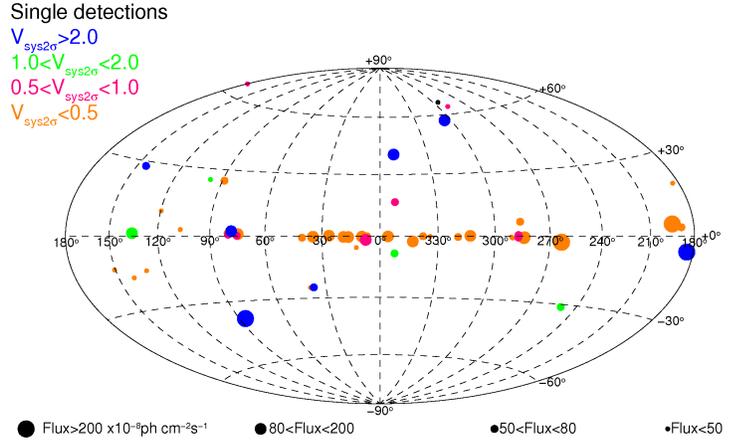}

\vskip -0.2truecm
\caption{ Aitoff projection of the \sounum\ distinct source positions detected on all pointed observations data (symbol colors indicate the variability parameter $V_{sys2\sigma}$ of the 53 sources for which we define with blue the highest and with orange the lowest value. \object{1AGL J1238+0406} has a single detection and so is marked in black. Marker sizes are proportional to flux). }

\label{fig:Fig1}
\end{figure}
\vskip 0.2truecm

We calculated a fourth variability index, $F_{sys}$, similar to that used in the Fermi catalogues \citep[][]{Abdo09,Abdo10}, defined as a $\chi^{2}$ again including a 10\% systematic error in the weights, to directly compare it with the Fermi 1FGL index. In our analysis each source has a different number of detections ($N_{det}$) and so this parameter is expected to behave as the $\chi^{2}_{n}$ distribution where the number of degrees of freedom (dof) is $n=N_{det} - 1$. For this reason we considered the {\em reduced} $F_{sys}$ ($Reduc.F_{sys} = F_{sys}/N_{dof}$), and compared it with the 1FGL variability index divided by 10 (i.e. the number of dof in 1FGL), and similarly with the 2FGL variability index divided by 23. We discuss in Section \ref{sec:discussconcl} a possible comparison with the 2FGL index as well, which is instead based on an ML ratio test statistic.

\section{Results: the catalogue and source variability}
\label{sec:results}

Applying our selection criteria, we obtained a sample of \sounum\ distinct sources (Fig. \ref{fig:Fig1}) with a total of 1209 OB detections.
Eight 1AGL sources (2 high latitude and 6 on the galactic plane) were not detected in this processing either because of the low OB exposures and/or due to specific complex Galactic regions. We show information about the mean flux of these sources in Table 3, as obtained in the 1AGL. 

We describe in the following the variability results (listed in Table \ref{tab:varres}) and the light curves of bright pulsars and of some prominent sources. We compared the light curves of sources with a known counterpart with those in the most recent Fermi catalogue \citep{Nolan12} for the time period common to both catalogues. The non-uniform exposures among the OBs together with the pointing strategy put strong constraints on the variability analysis. \\

\begin{table*}[tbh]
\begin{center}
\begin{tabular}{|llllcc|}

\hline
& & & &  &  \\
\multicolumn{1}{|l}{Source name} & 
\multicolumn{1}{c}{R.A. (J2000.0)} & 
\multicolumn{1}{c}{Dec. (J2000.0)} & 
\multicolumn{1}{c}{LII,BII} &
\multicolumn{1}{c}{Error Rad.} & 
\multicolumn{1}{c|}{Mean Flux\,$\pm$\,Error}  \\
\multicolumn{1}{|l}{        } & 
\multicolumn{1}{c}{  (hh mm ss) } & 
\multicolumn{1}{c}{  (dd mm ss) } & 
\multicolumn{1}{c}{ (deg,deg) } &
\multicolumn{1}{c}{ (deg) } & 
\multicolumn{1}{c|}{  ($\times 10^{-7}$ \pflux) }   \\[3 pt]
\hline
& & & &  &  \\[3 pt]
\object{1AGL J0657+4554} & 06 57 29.2 & +45 54 14.5 & 170.73 , 20.11 & 0.55 & 3.1 $\pm$ 0.6 \\[3 pt]

\object{1AGL J1043-5749} & 10 43 56.0 & -57 49 51.0 & 286.60 ,  0.94 & 0.68 & 2.2 $\pm$ 0.5 \\[3 pt]

\object{1AGL J1222+2851} & 12 22 39.7 & +28 51 02.3 & 196.09 , 83.42 & 0.74 & 3.8 $\pm$ 1.1 \\[3 pt]

\object{1AGL J1412-6150} & 14 12 06.1 & -61 49 32.5 & 312.30 , -0.43 & 0.44 & 4.3 $\pm$ 0.7  \\[3 pt]

\object{1AGL J1746-3017} & 17 46 01.5 & -30 17 23.7 & 358.89 , -0.78 & 0.68 & 6.6 $\pm$ 1.6 \\[3 pt]

\object{1AGL J1815-1732} & 18 15 29.7 & -17 32 27.1 &  13.29 , -0.28 & 0.65 & 5.2 $\pm$ 1.3 \\[3 pt]

\object{1AGL J1901+0430} & 19 01 20.8 & +04 29 38.5 &  38.06 , -0.15 & 0.58 & 4.5 $\pm$ 1.1 \\[3 pt]

\object{1AGL J1923+1404} & 19 22 53.7 & +14 03 45.2 &  49.00 , -0.42 & 0.64 & 6.0 $\pm$ 1.0 \\[3 pt]

\hline

\end{tabular}
\caption{
 1AGL sources not detected in this analysis. The information is taken from 1AGL, meaning these are first-year data. } 
\end{center}

\label{tab:1aglnot} 
\end{table*}

The source and detections table\footnote{Available on-line at\\{\it http://www.asdc.asi.it/agile1rcat}}, Table \ref{tab:list}, includes all the source flux measurements with $\sqrt{TS}$\,$\geq$\,2 obtained for each of the \sounum\ sources detected in this processing.
 In Table \ref{tab:list} we report the main parameters of the maximum $\sqrt{TS}$ detection for each source. These are the name, coordinates, the $\sqrt{TS}$, the E\,$>$\,100 \Mevd\ flux, the four variability indices described above and the number of detections, the confirmed counterparts and source class, if any, and other possible associations ordered according to the distance from the AGL source. We used the abbreviation {\em 1AGLR} for new sources and repositioned 1AGLs.
\begin{figure}[ht]
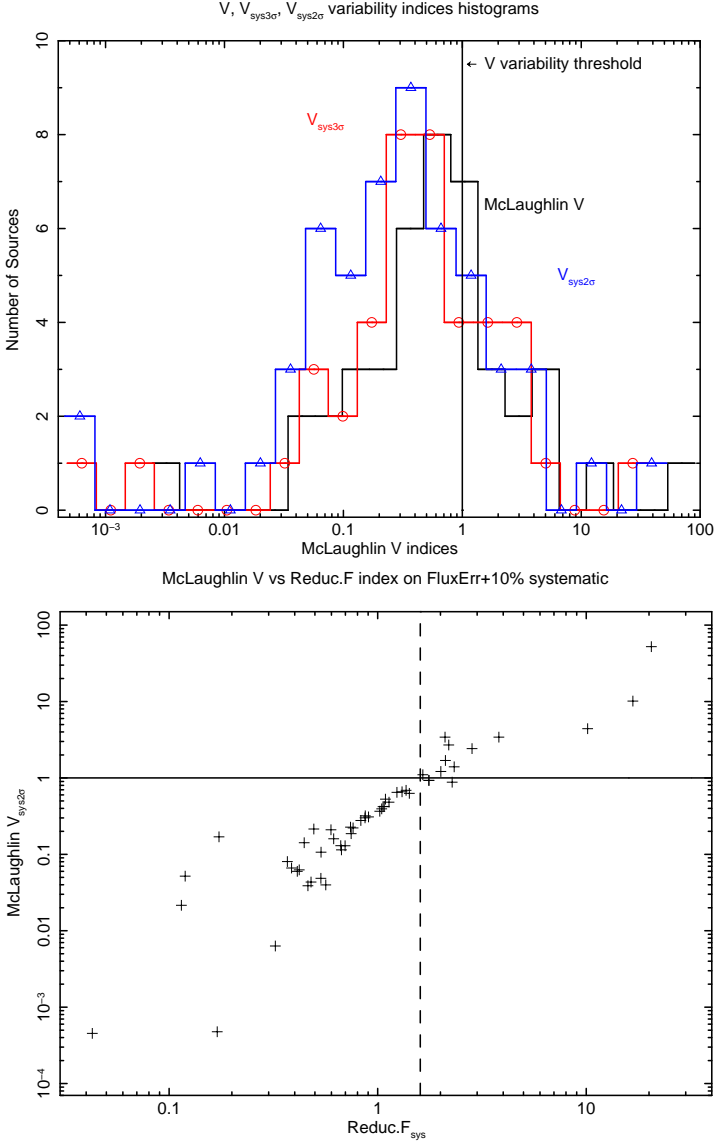

\begin{center}
\hskip -1.3truecm
\vbox{
\includegraphics[angle=-90,width=0.52\textwidth]{verrecchia_fig5.ps}

\includegraphics[angle=-90,width=0.51\textwidth]{verrecchia_fig6.ps}
}
\end{center}
\vskip -0.5truecm
\caption{
 Upper panel: histograms of the V (black unmarked line), V$_{sys2\sigma}$ (red line with circles), and V$_{sys3\sigma}$ (blue lines with triangles) parameters; lower panel: plot of V$_{sys2\sigma}$ vs the $Reduc.F_{sys}$. The vertical line in the upper plot is the threshold above which a source is variable (V\,$\geq$\,1) according to the V parameters. The horizontal line in the lower plot has the same meaning, while the vertical dashed line is the approximated extrapolation of a threshold for source variability for the $Reduc.F_{sys}$ parameter, taking into account the correlation of both parameters.
}
\label{fig:Fig7}
\end{figure}


Below this row a list with all the flux measurements in each OB is shown with the observation parameters MJD start/stop, the OB number, the OB $\sqrt{TS}$, the E\,$>$\,100 \Mevd\ flux or the ULs for 2\,$\leq$\,$\sqrt{TS}$\,$\leq$3, and the \mbox{off-axis} angle. The flux errors are statistical only, i.e. the results of our analysis. 
The association column is populated with sources from well-known catalogues that fall within error circles of radii given by the 95\% statistical plus the 0.1$^{\circ}$ systematic, linearly added. In particular, sources from the 1AGL were the first included when present, then sources from the BZCAT \citep{BZCAT}, from 3EG and 2FGL.
Associations from known radio \citep{OS,GB6}, TeV (TEVCAT\footnote{http://tevcat.uchicago.edu}), SNR \citep{Gr09}, PSR \citep{ATNF}, and HMXB \citep{HMXB} catalogues were included for some sources without a previously confirmed counterpart (TeV emitting PWNs associated with bright pulsars are not shown).
\vskip -0.4truecm

\subsection{Variability results}
\label{sec:varres}

 The $V$ index distributions are given in Fig.~\ref{fig:Fig7}. They show that V values peak in the non-variable--uncertain ranges for all indices and that the $V$ distribution is on average lower than the $V_{sys2\sigma}$ (average values for $V$ and $V_{sys2\sigma}$ are 3.6 and 1.8), because of the obviously higher number of detections used for the $V_{sys2\sigma}$.
The V$_{sys2\sigma}$ and $Reduc.F_{sys}$ gave similar results, as expected (Fig.~\ref{fig:Fig7}), with a correlation factor 0.84. We then used this correlation to estimate a variability threshold for $Reduc.F_{sys}$ extrapolated from the V$_{sys2\sigma}$ threshold, considering the approximate intersection of the V$_{sys2\sigma}$\,=\,1 line with the linear correlation plot. We estimated two possible linear correlation whose steepest and faintest slope included all points, obtaining a range of the $Reduc.F_{sys}$ values at the intersection of 1.55--1.65, so we chose as threshold $Reduc.F_{sys}\,\approx$\,1.6.\linebreak
Details on the variability results obtained for different source classes are given in Table \ref{tab:varres}, for which we consider only sources with confirmed counterparts, except for 9 high-latitude sources associated with well known blazars also included as AGNs.
The majority of variable sources are blazars according to all indices, the most variable one is 3C 454.3. 
The fraction of variable sources according to the $V_{sys2\sigma}$ index is 20\%, and a similar result, 22\%, is obtained according to $V_{sys3\sigma}$. 

\begin{table}[tbh]
\begin{center}
\begin{tabular}{|l l l|}
\hline
& & \\
\multicolumn{1}{|c}{Variability } & 
\multicolumn{1}{c}{Number of } & 
\multicolumn{1}{c|}{Source class: Number} \\[3 pt]
\multicolumn{1}{|c}{ } & 
\multicolumn{1}{c}{ Sources} & 
\multicolumn{1}{c|}{of variables } \\[3 pt]
\multicolumn{1}{|c}{Type } & 
\multicolumn{1}{c}{(V,V$_{sys3\sigma}$,V$_{sys2\sigma}$)} & 
\multicolumn{1}{c|}{ } \\[3 pt]
\hline
 & & \\[3 pt]
 Variable & 13, 12, 11 & AGN:7,8,8; PSR:1,0,0; \\[3 pt]
     &    & PSR$^{*}$:1,1,1;HMXB:1,1,1; \\[3 pt]
     &    & CWB:1,0,0;SNR:0,0,0; \\[3 pt]
     &    & Unidentified:2,2,1 \\[3 pt]
 Uncertain & 8, 5, 8 & AGN:2,1,3;PSR:3,2,1;  \\[3 pt]
     &    & HMXB:1,0,1;SNR:0,0,0;   \\[3 pt]
     &    & CWB:0,1,1;Unidentified:2,1,2   \\[3 pt]
 Non-variable & 21, 25, 34 & AGN:7,7,6; PSR:4,6,8; \\[3 pt]
     &    & HMXB:0,1,0;SNR:2,2,2;  \\[3 pt]
     &    & CWB:0,0,0;Unidentified:8,9,18  \\[3 pt]
\hline
\end{tabular}
\end{center}

\caption{Variability analysis results according to the V, V$_{sys3\sigma}$, and V$_{sys2\sigma}$ indices, divided in to six main source classes. The last column lists the number of sources of each variability type found in each source class according to each index, in the same order as in the second column. The PSR$^{*}$ source is the Crab pulsar + nebula.} 
\label{tab:varres} 
\vskip -0.2truecm
\end{table}

However, the $F_{sys}$ parameter and consequently V (which is based on the calculation of the $\chi^{2}$), depend on the ratio between mean flux and the standard deviation, so it is expected to be more likely to classify faint sources (flux $<$3.0\,$\times$\,10$^{-7}$\pflux) or those near the Galactic plane as non-variable. Most of the unclassified sources are non-variable, but about half of the unclassified have an unconfirmed association with 2FGL pulsars.

\begin{figure}[ht]
\begin{center}
\hskip -0.8truecm
\vskip -0.3truecm

\includegraphics[angle=-90,width=0.51\textwidth]{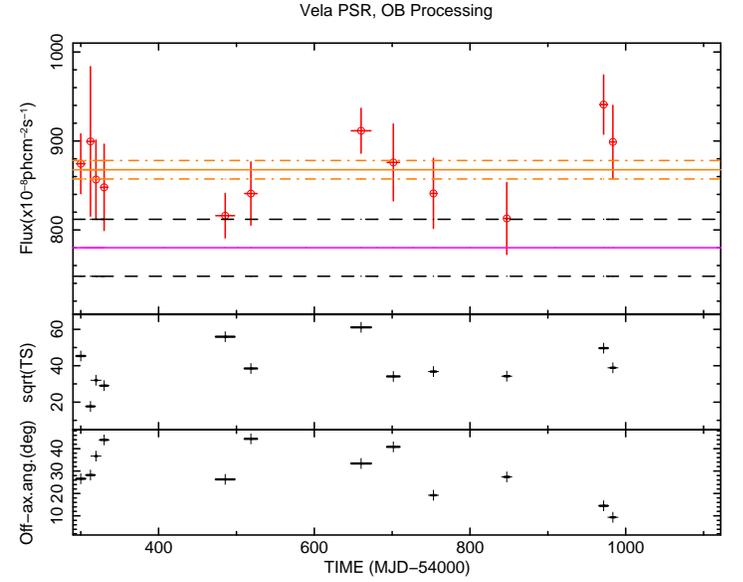}
\end{center}
\caption{Vela pulsar E\,$>$\,100\,\Mevd\ light curve in 10$^{-8}$ \pflux\ (upper panel), the $\sqrt{TS}$ (middle panel), and the \mbox{off-axis} angle (in degrees) as a function of time in MJD (lower panel). The 1AGL flux value and the 1\,$\sigma$ error levels are shown as a magenta line and black dashed lines, while the weighted mean value of all OBs and its 1\,$\sigma$ error levels are shown in orange.
}
\label{fig:Fig5}
\end{figure}

\subsection{Light curves of bright pulsars }

\begin{figure}[ht]
\begin{center}

\hskip -0.3truecm
\vbox{

\includegraphics[angle=-90,width=0.52\textwidth]{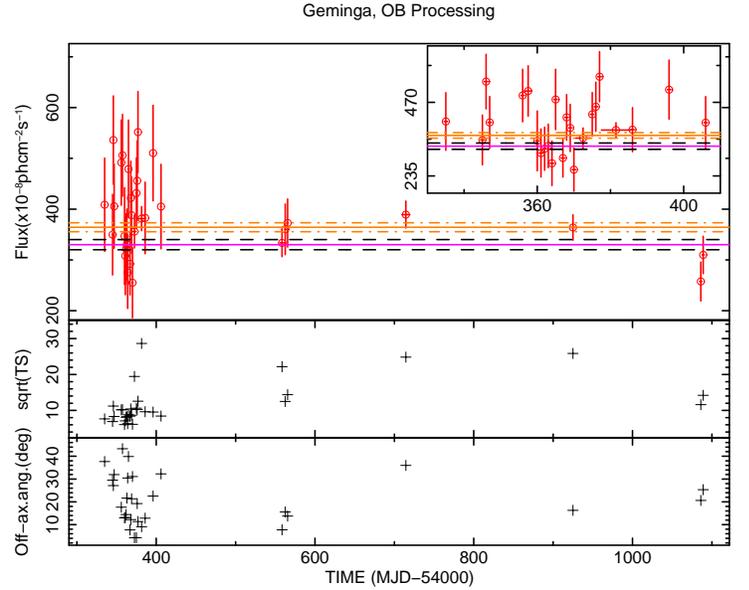}

}
\end{center}
\vskip -0.4truecm
\caption{Geminga pulsar E\,$>$\,100\,\Mevd\ light curve in 10$^{-8}$ \pflux\ (upper panel), the $\sqrt{TS}$ (middle panel), and the \mbox{off-axis} angle (in degrees) as a function of time in MJD (lower panel) for detections with an \mbox{off-axis} angle lower than 50$^{\circ}$ and exposures greater than $9\times 10^{6}$ cm$^{2}$s. The 1AGL flux value and the 1\,$\sigma$ error levels are shown as magenta line and black dashed lines, while the weighted mean flux and its 1\,$\sigma$ error levels are shown in orange (solid and dash-dotted lines).}

\label{fig:Fig6}
\end{figure}

\begin{figure}[ht]
\begin{center}
\hskip -1.3truecm
\vbox{

\includegraphics[angle=-90,width=0.52\textwidth]{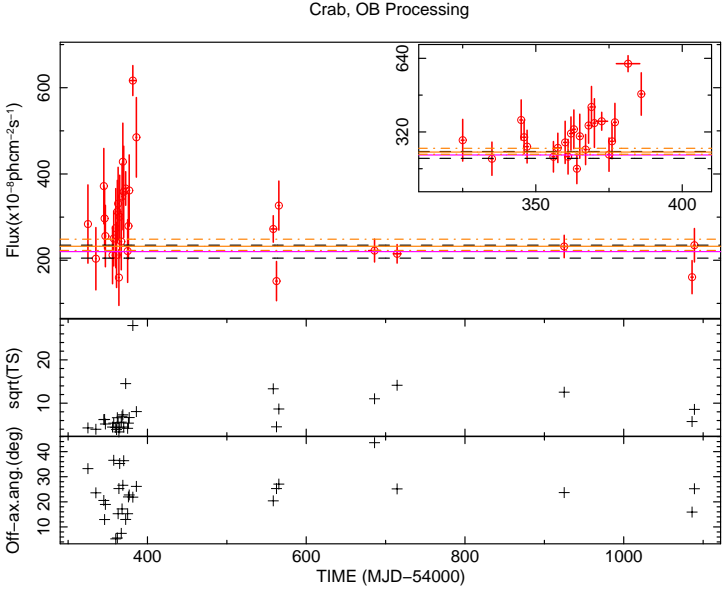}
}
\end{center}
\vskip -0.4truecm
\caption{Crab E\,$>$\,100\,\Mevd\ light curve (upper panel), the $\sqrt{TS}$ (middle panel), and the \mbox{off-axis} angle (in degrees) as a function of time (lower panel) for detections with an \mbox{off-axis} angle lower than 50$^{\circ}$ and exposures greater than $9\times 10^{6}$ cm$^{2}$s. The 1AGL flux values and the 1\,$\sigma$ error levels are shown as magenta line and black dashed lines, while the weighted mean value of all OB fluxes and its 1\,$\sigma$ error levels are shown in orange.
 The 2007 Crab flare episode (MJD=54381.5) is its most significant detection at a flux of 62.0$\pm$\,7.0\,$\times$\,10$^{-7}$\pflux. Large error bars at MJD\,$\leq$\,54420 are mostly obtained with 1-3 day exposures.}

\vskip -0.2truecm
\label{fig:Fig6b}
\end{figure}

 The light curves of the Vela and Geminga pulsars are shown in Figs. \ref{fig:Fig5} and \ref{fig:Fig6}, and that of the Crab pulsar + nebula in Fig.~\ref{fig:Fig6b}. Vela and Geminga are the brightest among the non-variable sources. In all figures the mean \mbox{off-axis} angles and the $\sqrt{TS}$ in each OB are shown as well.

The weighted mean of the OB Vela fluxes is 86.8\,$\pm$\,1.0 $\times$\,10$^{-7}$\pflux\ while the value from the updated deep maps is 88.2\,$\pm$\,0.9\,$\times$\,10$^{-7}$\pflux, which is $\sim$\,3.2\,$\sigma$ higher than the mean value from the first AGILE catalogue, 78.0$\pm$\,3.2\,$\times$\,10$^{-7}$\pflux, according to the 1AGL error. We did not overplot the new mean value estimated from deep maps because the variability study is based on the mean of the OB fluxes.
 This new value, which should be considered as the new reference one, is higher than the previous value primarily because of the new instrument response functions, which compensate for the effect of energy dispersion, as well as the new filter calibrated up to a high \mbox{off-axis} angle.
We obtained $V_{sys2\sigma}$\,=\,0 and $F_{sys}$\,=\,1.87, with 13 dof and expect a constant source to produce a value of $F_{sys}$\,$\leq$\,2.1 at 99\% c.l..

A particular case is that of the Crab Nebula, which has recently been discovered to be a variable $\gamma$-ray source \citep[see Tavani et al. 2011a, 20011b,][]{Abdo11}. In Fig.~\ref{fig:Fig6}, first panel, we plot the Geminga light curve and its weighted mean OB flux value, 36.4$\pm$\,0.9\,$\times$\,10$^{-7}$\pflux, is shown. In Fig.~\ref{fig:Fig6b} the Crab light curve shows the October 2007 flaring episode (MJD=54381.5, OB 4200), discussed in Tavani et al. (2011b), at a higher flux than for the 2010 flare. The Crab (Nebula + PSR) weighted mean OB flux value, 23.2$\pm$\,0.9\,$\times$\,10$^{-7}$\pflux, was calculated excluding seven OB during the 2007 flare (in the interval 54368--54396).
 We obtained V$_{sys2\sigma}$\,=\,0.4 for Geminga and V$_{sys2\sigma}$\,=\,3.4 for Crab. If we remove the 2007 flare peak from the Crab light curve we obtain V$_{sys2\sigma}$=\,0.8, while if we remove all seven OB that cover the flare period that was excluded for the mean value estimate, we obtain V$_{sys2\sigma}$=\,0.1, a non-variable value.

\subsection{Long-term light curves of some prominent sources}
\label{sec:specific}

\begin{figure}[ht]
\begin{center}


\includegraphics[angle=-90,width=0.50\textwidth]{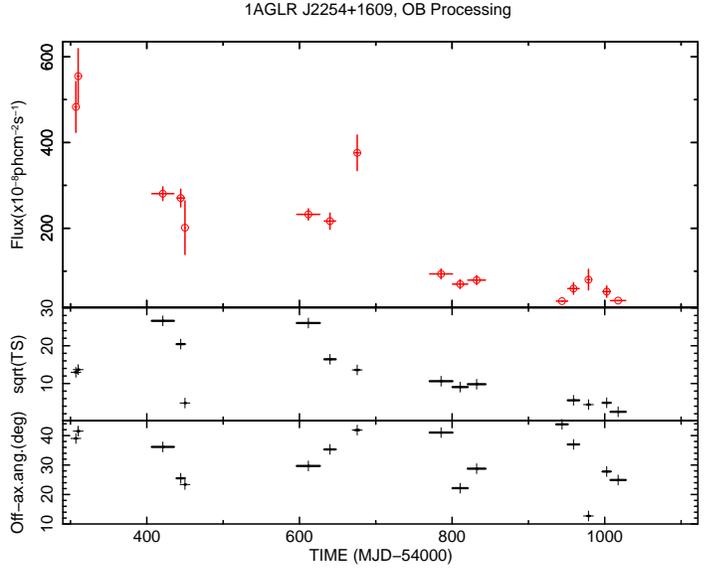}
\end{center}
\caption{3C 454.3 E\,$>$\,100\,\Mevd\ light curve in 10$^{-8}$ \pflux\ (upper panel), the $\sqrt{TS}$ (middle panel), and \mbox{off-axis} angle (in degrees) as a function of time in MJD (lower panel).
}
\vskip -0.1truecm
\label{fig:Fig8}
\end{figure}

\begin{figure}[ht]
\vskip -0.2truecm
\rotatebox[]{-90}{
\includegraphics[width=0.42\textwidth]{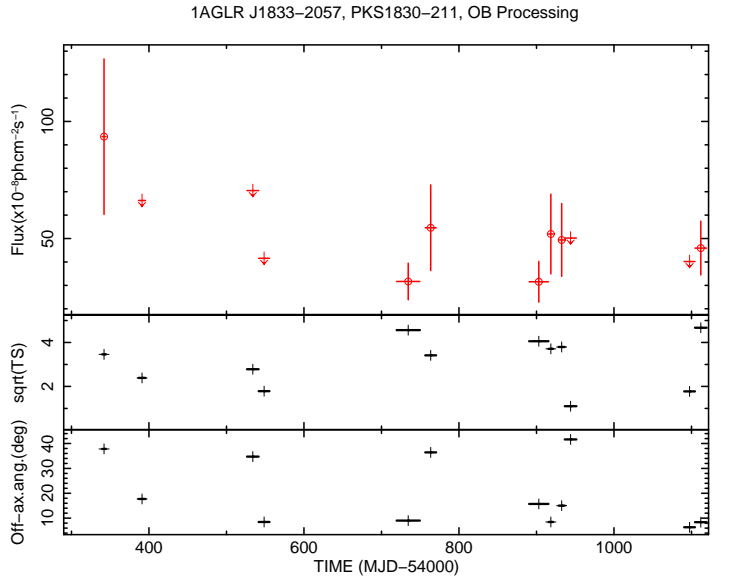}
}
\vskip -0.7truecm
\caption{\object{1AGLR J1833-2057}/PKS 1830-211 E\,$>$\,100\,\Mevd\ light curve in 10$^{-8}$ \pflux\ (upper panel), the $\sqrt{TS}$ (middle panel), and the \mbox{off-axis} angle (in degrees) as a function of time in MJD (lower panel).
}
\label{fig:PKS1830}
\end{figure}

We report here four long-term light curves, in the E\,$>$100\,\Mevd\ band, obtained for sources classified as variable, non-variable and a particular case of an uncertain one. We show in Figs.~\ref{fig:Fig8} and \ref{fig:PKS1830} the results for the two known blazars 3C 454.3 and PKS 1830-211. Light curves are shown including source detections with flux errors at 1$\sigma$, not including any systematic error.

 Fig.~\ref{fig:Fig8} shows the strong variability in the 3C454.3 flux even at the OB time scale, and in particular the decreasing trend reported in Vercellone et al. (2010). This source has the strongest variability index V$_{sys2\sigma}$\,=\,52.3. This confirms what was expected in Vercellone et al. (2004), a study of EGRET blazar activity and duty-cycle evaluation, which was based on a differently defined variability index.

\object{1AGLR J1833-2057}, associated to \object{PKS 1830-211}, is instead one of the fifteen new sources not included in 1AGL. Its light curve shows a detection in the period of the October 2009 flare \citep{stri1830}, with a flux of (3.2\,$\pm$\,0.8)\,$\times$\,10$^{-7}$\pflux, compatible with the one-week average flux (4.0\,$\times$\,10$^{-7}$\pflux) reported in Donnarumma et al. (2011), based on data of the OB 8300 only, and shown to be three times weaker than the peak flare flux.
This source is classified as non-variable according to all parameters, however, because of the flux averaging over the OB time scale.
Other detections have been obtained, the most significant one in September 2008. This detection showed some source activity one year before the October 2009 flare, with a time scale between the two events similar to that between the 2009 and the following 2010 flare, which was also detected by Fermi (see discussion in Donnarumma et al. 2011).

\begin{figure}[ht]
\vskip -0.3truecm
\hskip -0.5truecm
\rotatebox[]{-90}{
\includegraphics[width=0.42\textwidth]{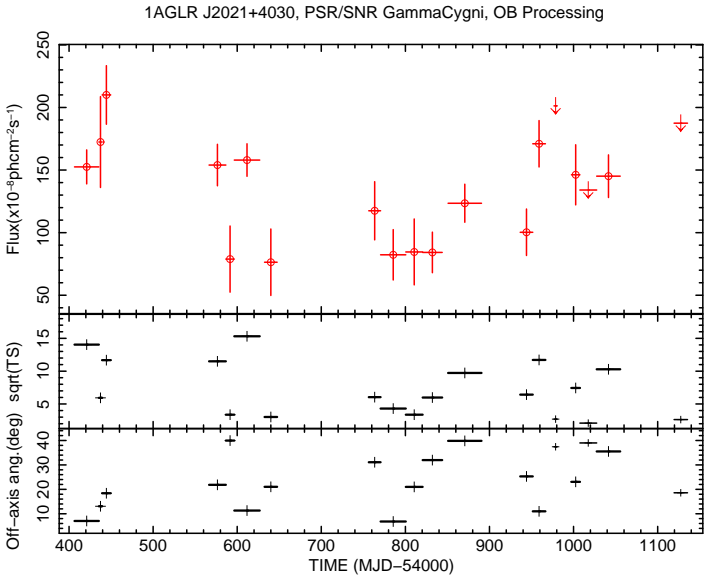}
}
\vskip -0.5truecm
\caption{\object{1AGLR J2021+4030} E\,$>$\,100\,\Mevd\ light curve in 10$^{-8}$ \pflux\ (upper panel), the $\sqrt{TS}$ (middle panel), and the \mbox{off-axis} angle (in degrees) as a function of time in MJD (lower panel).
}
\label{fig:GammaCyg}
\end{figure}

In Fig.~\ref{fig:GammaCyg} we show the \object{1AGLR J2021+4030} light curve, a source that falls in the $\gamma$ Cygni SNR region.
 The 1AGLR J2021+4030 variability on six-day time scale on pointing data (up to August 2009) was extensively discussed in Chen et al. (2011a), where values of 3.88 and 2.18 in E\,$>$\,100\,\Mev\ were found for $V$ and $V_{sys2\sigma}$, respectively, while low variability was detected at E\,$>$\,400\Mevd. The corresponding values in this paper are 5.9 and 2.70 and the light curve is compatible with that therein. 1AGLR J2021+4030 field is complex and there could be contribution to the variability from other $\gamma$-ray sources, that are possibly transient, in addition to LAT \object{PSR J2021+4026} \citep[see ][for more details]{chen11}.

\begin{figure}[ht]
\vskip -0.2truecm
\hskip -0.5truecm
\rotatebox[]{-90}{
\includegraphics[width=0.41\textwidth]{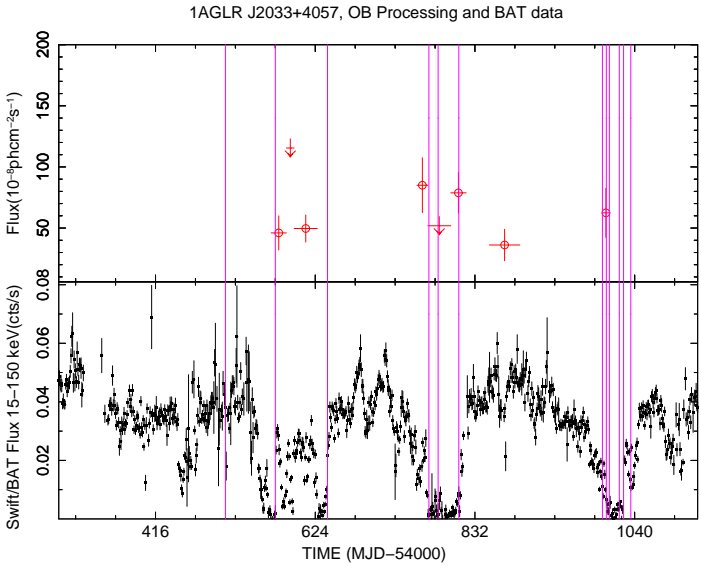}
}
\vskip -0.7truecm
\caption{\object{1AGLR J2033+4057}/Cyg X-3 E\,$>$\,100\,\Mevd\ light curve (upper panel) compared with the Swift/BAT 15\,$<$\,E\,$<$\,150\,\kev\ daily light curve. The plot shows six detections with $\sqrt{TS}$\,$>$3.
The 1--2 days $\gamma$-ray flare times described in Tavani et al. (2009c), Bulgarelli et al. (2012a), and Piano et al. (2012) are marked with vertical magenta lines. For some lines there is no corresponding UL when they have $\sqrt{TS}$\,$<$2.
}
\label{fig:CygX-3}
\end{figure}

Finally, we show in Fig.~\ref{fig:CygX-3} the \object{1AGLR J2033+4057}/ Cyg~X-3 light curve together with the 15--150\,\kev\ daily light curve from Swift/BAT (Barthelmy et al. 2005). This plot is similar to those reported in  Tavani et al. (2009c), Bulgarelli et al. (2012a), and Piano et al. (2012).
Here we show the 1--2-day $\gamma$-ray flare times in the Pointing mode interval described there. Our detections correspond to the reported activity periods, and moreover a flux UL at $\sqrt{TS}$\,$=$\,2.7 (MJD=54591.5) is compatible with a short {\em dip} in the BAT hard X-ray light curve. We performed an additional ML analysis on a three-day time scale that found an additional low-significance ($\sqrt{TS}$\,=\,4) short time scale $\gamma$-ray emission enhancement not included in Piano et al. (2012), which in fact reported the $\gamma$-ray activity on the 1--2-day time scale. 
 Again our flux variations are averaged on longer time intervals, so that this source is only classified as uncertain in this analysis.

\section{Discussion and conclusions}
\label{sec:discussconcl}

We reported the results of a variability analysis in the E\,$>$\,100\,\Mev\ band of a catalogue of \sounum\ sources obtained as an update of the 1AGL using the complete AGILE pointing mode OB archive. The new source list includes seven previously undetected sources with AGN counterparts and eight at low latitudes, while eight 1AGL sources were not detected in this analysis, two of which are at high latitude (\object{1AGL J0657+4554} and \object{1AGL J1223+2851}/WComae) and six are on the Galactic plane (\object{1AGL J1044-5750}, \object{1AGL J1412-6150}, \object{1AGL J1746-3017}, \object{1AGL J1815-1732}, \object{1AGL J1901+0430}, \object{1AGL J1923+1404}).

We carried out the variability study by applying an ML analysis in each single OB without implementing any blind search for transient sources, among OBs with different exposure times.

 This ML variability analysis, as discussed in Section \ref{sec:detselvar}, is known to be inadequate for the variability indices used \citep{Nolan03}; however, the revised indices allowed us to distinguish variable from non-variable sources. All four variability indices defined in our analysis show that the majority of variable sources are blazars, consistent with the $\gamma$-ray source variability studies included in the first (1FGL) and second (2FGL) Fermi source catalogues. 

\begin{figure}[ht]
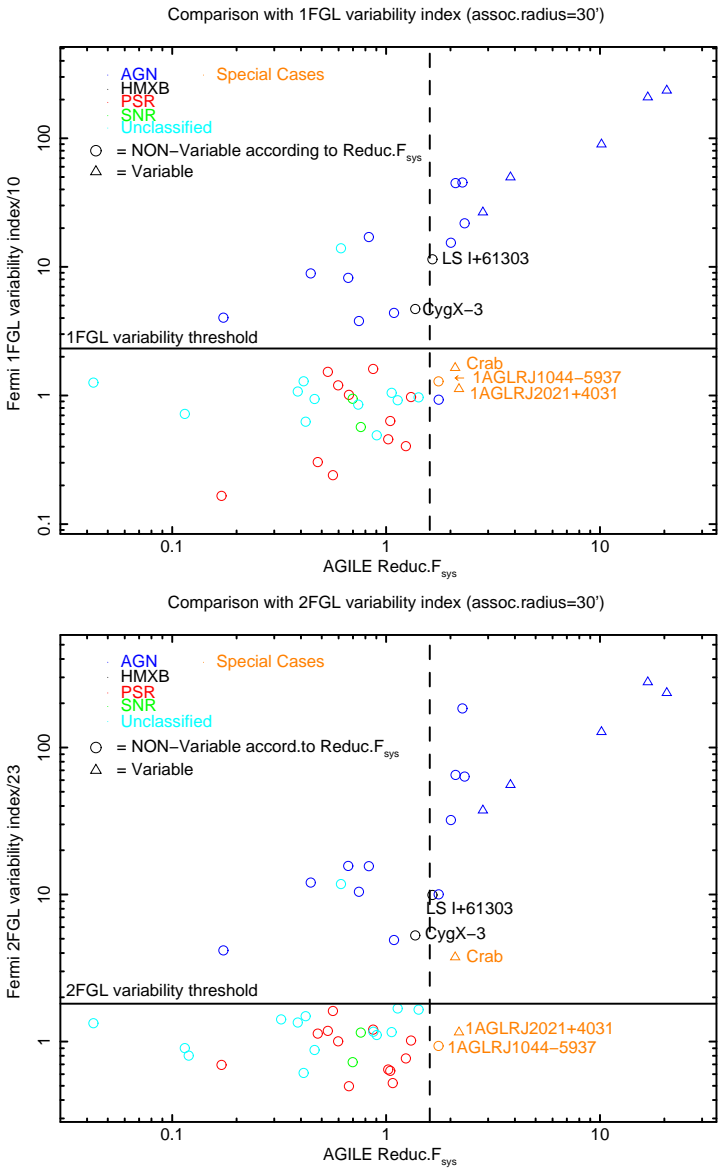

\begin{center}
\vskip -0.2truecm
\hskip -1.2truecm
\vbox{
\includegraphics[angle=-90,width=0.515\textwidth]{verrecchia_fig14.ps}


\includegraphics[angle=-90,width=0.515\textwidth]{verrecchia_fig15.ps}
}
\end{center}
\vskip -0.4truecm
\caption{
 Upper plot: Fermi 1FGL variability index divided by 10 vs $Reduc.F_{sys}$ for all detections associated with the 1FGL sources with a cross-matching with radius 30$\arcmin$, where colours correspond to source classes, blue for AGNs, black for HMXBs, red for PSRs, green for SNRs, light-blue for the unclassified ones, and orange for the three {\em Special Cases} Crab nebula + pulsar, \object{1AGLR J2021+4030} ($\gamma$ Cygni region) and \object{1AGLR J1044-5944} (Eta Carinae). Markers correspond to AGILE variability type according to $Reduc.F_{sys}$, i.e. circles for non-variable and triangles for variable. The lower plot is obtained with association to the 2FGL catalogue with the same radius and shows the 2FGL index divided by 23. The black horizontal lines are the 1FGL and 2FGL variability thresholds, 2.32 and 1.81, while the dashed vertical lines correspond to the approximately extrapolated threshold (1.6) for the AGILE variability $Reduc.F_{sys}$ parameter obtained from the correlation with the $V_{sys2\sigma}$ parameter.
}
\label{fig:FigLast}
\end{figure}
A comparison with the 1FGL and 2FGL variability indices is shown in Fig.~\ref{fig:FigLast}. The index from the 1FGL is a $\chi^{2}$ statistic calculated in the same manner as $F_{sys}$, while the 2FGL statistic is an ML ratio test statistic, defined in Nolan et al. (2012), which is distributed according to the $\chi^{2}_{23}$ distribution in the null hypothesis of a constant source.

We associated our source list with those of 1FGL and 2FGL by a simple spatial cross-matching with radius 30$\arcmin$, yielding 46 and 49 sources in common, respectively. We used the $Reduc.F_{sys}$ parameter because each source is detected in a different number of OBs (see Section \ref{sec:detselvar}).

For each source we defined a specific threshold for flux variability at 99\% c.l. \citep[as in ][]{Abdo10,Nolan12} using the appropriate $\chi^{2}_{n}$ distribution for $n$ dof. Thresholds range from 1.67 for a source with 31 detections to 6.2 for one with two detections, while the 1FGL and 2FGL thresholds are  2.32 and 1.81, respectively. Twelve sources out of \sounum\ have fewer than five detections. Sources with a $Reduc.F_{sys}$ higher than their own threshold ($Reduc.F_{sys}$\,$>$\,$\chi^{2}_{thres}/N_{dof}$) are flagged as variable in the plot. We found that all of the AGILE variable sources, seven sources, are also variable in 1FGL except for two, the Crab and 1AGLR J2021+4030, while in 2FGL only 1AGLR J2021+4030 remains non-variable. The Crab case is clear: the AGILE pointing dataset included the very intense 2007 flare, with a peak flux three times the 1AGL mean flux, whose evolution is shown in various consecutive short-duration OBs (see Table \ref{tab:pointings}), while fewer pointings are available from 2008 on. 1FGL included a flare affecting only the last of the 11 bins in the catalogue, while the 2FGL results show a much higher variability than for a lower mean flux with a smaller error.
As discussed in Section \ref{sec:specific}, the variable emission from 1AGLR J2021+4030 might be generated by the contribution from other sources apart from the pulsar.

Sources in the plots are divided into five main classes: AGN, PSRs, HMXBs, SNRs, and unclassified sources, while three particular cases are labelled separately (Crab PSR + nebula, 1AGLR J2021+4030 and \object{1AGLR J1044-5944}). Only sources with confirmed (based on 1AGL and later AGILE results) counterparts were included in the specific classes, except for 9 high-latitude sources associated to well known blazar counterparts included as AGNs. 

A good correlation with the 1FGL parameter is obtained for AGN whose diffuse $\gamma$-ray background is low, with a correlation factor of 0.98, which is expected for this class, taking into account their broad flux distribution and that at low fluxes ($\leq$\,3.0\,$\times$\,10$^{-7}$\,\pflux\ for a typical AGN variability of 1-3 days) AGILE cannot detect variability. Lower correlation factors are obtained for PSRs and unclassified (0.01 and 0.05). The correlation factor considering all sources together is 0.98. The PSRs are all non-variable except for the two particular cases discussed above.

Lower values were also obtained for the correlation of $Reduc.F_{sys}$ with the 2FGL index (0.88 for AGN, -0.28, 0.08 for PSRs and unclassified, while the value including all sources is 0.89), for AGN due also to 3C 273, whose variability index increased by a factor 4 from 1FGL to 2FGL (excluding it, the correlation factor for AGN is 0.97) and to \object{1AGLR J1625-2531} which became variable in 2FGL with an increase of a factor 10 in the variability index.
We note the presence of an unclassified source within the AGN group in both plots. This is \object{1AGLR J2016+3644}, a source within the Cygnus region that is non-variable according to $Reduc.F_{sys}$ (and also $V_{sys2\sigma}=0.2$), whose nearest Fermi sources are the unclassified \object{1FGL J2015.7+3708} (at 25$\arcmin$), \object{2FGL J2015.6+3709}/MG2 J201534+3710 (at 27$\arcmin$), classified as a blazar of uncertain type, \object{2FGL J2018.0+3626} (unassociated, at 27$\arcmin$) and the SNR \object{G74.9+1.2} (28$\arcmin$). All sources are slightly outside the 1AGLR J2016+3644 error circle, however.
The comparison with the vertical dashed lines, the threshold at 1.6, shows that some sources marked with circles could be variable according to $V_{sys2\sigma}$, as confirmed by the exact numbers in Table \ref{tab:varres}. For instance, eight AGNs are variables according to $V_{sys2\sigma}$ and ten blue points are rightward of the dashed line, while there are only five triangles.

The inhomogeneous exposure times among the OBs diminish the capability of studying the source variability in our data analysis; for this reason a similar analysis on the one-week time scale on the same pointing data archive would be useful to better classify the source variability, and will be reported in a future work.

\begin{acknowledgements}
The AGILE Mission is funded by the Italian Space Agency (ASI) with
scientific and programmatic participation by the Italian Institute
of Astrophysics (INAF) and the Italian Institute of Nuclear
Physics (INFN). Research partially supported through the ASI grants I/089/06/2, 
I/042/10/0 and I/028/12/0. The authors thank the anonymous referee for the useful suggestions.

\end{acknowledgements}



\bibliography{verrecchia_proofar}


\hrule
\begin{list}{}{}
\topsep
\itemsep
\item $^{1}$ ASI Science Data Center (ASDC), via del Politecnico snc, 00133 Roma, Italy.
\item $^{2}$ INAF-Osservatorio Astronomico di Roma, Via di Frascati 33, 00040 Monte Porzio Catone, Italy.
\item $^{3}$ INAF/IASF Milano, via E. Bassini 15, 20133 Milano, Italy.
\item $^{4}$ INAF/IASF Bologna, via Gobetti 101, 40129 Bologna, Italy.
\item $^{5}$ INAF/IAPS, via del Fosso del Cavaliere 100, 00133 Roma, Italy.
\item $^{6}$ Dip.di Fisica, Universit\`a ``Tor Vergata'', via della Ricerca Scientifica 1, 00133 Roma, Italy.
\item $^{7}$ INFN-Roma ``Tor Vergata'', via della Ricerca Scientifica 1, 00133 Roma, Italy.
\item $^{8}$ Consorzio Interuniversitario Fisica Spaziale (CIFS), villa Gualino - v.le Settimio Severo 63, 10133 Torino, Italy.
\item $^{9}$ Agenzia Spaziale Italiana, viale Liegi 26, 00198 Roma, Italy.
\item $^{10}$ INAF-Osservatorio Astronomico di Cagliari, loc. Poggio dei Pini, strada 54, 09012, Capoterra (CA), Italy.
\item $^{11}$ INAF-IASF Palermo, Via Ugo La Malfa 153, 90146 Palermo, Italy.
\item $^{12}$ Dip. Fisica, Universit\`a di Trieste, via A. Valerio 2, 34127 Trieste, Italy.
\item $^{13}$ INFN Trieste, Padriciano 99, 34012 Trieste, Italy.
\item $^{14}$ INFN Pavia, via A. Bassi 6, 27100 Pavia, Italy.
\item $^{15}$ Dip. di Fisica, Universit\`a di Torino, via P. Giuria 15, 10126 Torino, Italy.
\item $^{16}$ ENEA Bologna, via don Fiammelli 2, 40128 Bologna, Italy.
\item $^{17}$ INFN Roma 1, p.le Aldo Moro 2, 00185 Roma, Italy.
\item $^{18}$ Dip. Fisica, Universit\`a ''La Sapienza'', p.le Aldo Moro 2, 00185 Roma, Italy.
\item $^{19}$ ENEA Frascati, via Enrico Fermi 45, 00044 Frascati (RM), Italy.
\item $^{20}$ CNR, IMIP, Area Ricerca Montelibretti (RM), Italy.
\item $^{21}$ Dip. Fisica, Universit\`a dell'Insubria, Via Valleggio 11, 22100 Como, Italy.
\item $^{22}$ ASTRON, the Netherlands Institute for Radio Astronomy, Postbus 2, 7990 AA, Dwingeloo, The Netherlands
\item $^{23}$ Wits University, 1 Jan Smuts Avenue Braamfontein 2000, Johannesburg, South Africa.
\end{list}

\clearpage
\newpage

\markright{F. Verrecchia \etal: An updated list of AGILE bright $\gamma$--ray sources and their variability in pointing mode}
\hspace{15cm}

\vspace{8.0cm}

\centering \hfill \hfill \hfill \hfill \hfill \hfill \hfill \hskip 2.0cm \hbox{\large {Supplementary Table (Table 5)}}

\clearpage
\onecolumn
 %
 %



\begin{landscape}
\setcounter{page}{14}
\markboth{}{}

\scriptsize

\setcounter{table}{4}

\LTcapwidth=10.2in

\begin{longtable}{|llllllllllllll|}

\caption{The variability results for the revised AGILE bright $\gamma$--ray sources list over the pointing-mode dataset. Column descriptions are provided at the end of the table.}  \\
\hline\hline
& & &     & & & &   &  & &   & & & \\ 
 AGILE name &  R.A. (J2000.0) & Dec. (J2000.0) & LII,BII  ($^{a}$Error Rad.) & $^{b}$TIME &   $\sqrt{TS}$ & $^{d}$Flux\,$\pm$\,Error & $^{e}$1AGL  & $^{f}$V , $^{g}$V$_{sys3\sigma}$,V$_{sys2\sigma}$ & $^{h}F_{sys}$ & $^{i}$Num. & Class & Confirmed  & 1AGL\&Possible \\
 & (hh mm ss)& (dd mm ss)  &   ~~($^\circ$,$^\circ$)  ~~~~ ~~~  ($^\circ$) & year.doy  & &   & Flux   &   Mc L. indices     & &  det. & & Counterp. & Association   \\

\hline
& & &     & & & &   &  & &   & & & \\ 
\endfirsthead

\caption{Continued. Column descriptions are provided at the end of the table.}\\
\hline\hline
& & &     & & & &   &  & &   & & & \\ 
 AGILE name &  R.A. (J2000.0) & Dec. (J2000.0) & LII,BII  ($^{a}$Error Rad.) & $^{b}$TIME &   $\sqrt{TS}$ & $^{d}$Flux\,$\pm$\,Error & $^{e}$1AGL  & $^{f}$V , $^{g}$V$_{sys3\sigma}$,V$_{sys2\sigma}$ & $^{h}F_{sys}$ & $^{i}$Num. & Class & Confirmed  & 1AGL\&Possible \\
 & (hh mm ss)& (dd mm ss)  &   ~~($^\circ$,$^\circ$)  ~~~~ ~~~  ($^\circ$) & year.doy  & &   & Flux   &   Mc L. indices     & &  det. & & Counterp. & Association   \\

\hline
& & &     & & & &   &  & &   & & & \\
\endhead

\hline
\hline
\endfoot

\hline
\endlastfoot
1AGLR J0007+7307 &  00 07 03.6 &  73 06 51.6 &  119.67 ,  10.52 ~ (0.08) &  2009.039 &  9.1 &  32.3$\pm$4.5 &  23.0 &  0.1 ~,  0.05/0.05  &             6.41 &   13 & Pulsar  &  PSRJ0007+7303 &  1AGLJ0006+7311 \\ 
& & &     & & & &   &  & &   & & & 3EGJ0010+7309 \\
& & &     & & & &   &  & &   & & & 2FGLJ0007.0+7303 \\
\hline
  & Start MJD & End MJD  & OB Num. & $^{c}$MeanOB & $\sqrt{TS}$ &
   $^{c}$Flux $\pm$ Error  & Off-axis &     & & & & &   \\
   & &   &        &  Exp.  & &   /U.L. & angle ($^{\circ}$)     & & & & & &  \\
\hline
  & 54308.5   & 54311.5   &  1150 &   21.8 &  4.55 &  83.7$\pm$25.0    &   35.4 &  & & & & &   \\ 
  & 54347.5   & 54355.5   &  2300 &   59.7 &  3.68 &  30.4$\pm$10.3    &   17.3 &  & & & & &   \\ 

  & 54439.4   & 54449.5   &  4910 &   74.4 &  3.89 &  35.2$\pm$11.3    &   38.3 &  & & & & &   \\
\end{longtable}

\normalsize
\flushleft
\begin{minipage}{240mm}
$^{a}$ Positional error circle radius at 95\% c.l., statistical error only. 
The AGILE team recommends adding a systematic error of $\pm 0.1$ degrees linearly. \\
$^{b}$ Start time in year.DoY.  \\
$^{c}$ Mean value of the exposure map relative to the OB map used for each source analysis in cm$^{2}$ Ms.\\
$^{d}$ E$>$100 {\rm MeV} flux and its $1 \sigma$ statistical error in $10^{-8}$ ph cm$^{-2}$ s$^{-1}$ units. 
The AGILE team recommends adding a systematic error of $10\%$ to the statistical error. \\
$^{e}$ E$>$100 {\rm MeV} flux reported in 1AGL in $10^{-8}$ ph cm$^{-2}$ s$^{-1}$ units.  \\
$^{f}$ McLaughlin V variability index (McLaughlin et al. 1996).  \\
$^{g}$ McLaughlin V variability indices calculated including a 10\% systematic error and for the last value (after ``/'') also including detections at 2$\leq\,\sqrt{TS}$\,$<$\,3.  \\
$^{h}$ A modified $\chi^{2}$ including a 10\% systematic error.  \\
$^{i}$ Number of detections with $\sqrt{TS}$\,$\geq$2.  \\
$^{*}$ Pulsar + nebula.  \\
$^{**}$ $\gamma$--ray source positionally consistent with the pulsar 2FGLJ1836.2+5926.  \\

\end{minipage}

\end{landscape}
\twocolumn

\clearpage


\end{document}